# Anomalous evolution of the magnetocaloric effect in dilute triangular Ising antiferromagnets Tb$_{1-x}$Y$_x$(HCO$_2$)$_3$

Mario Falsaperna[1], Johnathan M. Bulled[2], Gavin B. G. Stenning[3], Andrew L. Goodwin[2], Paul J. Saines[1*].

[1]School of Chemistry and Forensic Science, Ingram Building, University of Kent, Canterbury, CT2 7NH, UK
[2]Inorganic Chemistry Laboratory, South Parks Road, Oxford OX1 3QR, UK
[3]ISIS Neutron and Muon Source, STFC Rutherford Appleton Laboratory, Chilton, Didcot, OX11 0QX
*Corresponding Author Email Address: P.Saines@kent.ac.uk

**Abstract**

We investigate the effects of diamagnetic doping in the solid-solution series Tb$_{1-x}$Y$_x$(HCO$_2$)$_3$, in which the parent Tb(HCO$_2$)$_3$ phase has previously been shown to host a combination of frustrated and quasi-1D physics, giving rise to a triangular Ising antiferromagnetic ground state that lacks long range 3D order. Heat capacity measurements show three key features: (i) a low temperature Schottky anomaly is observed, which is constant as a function of $x$; (ii) the transition temperature and associated entropy change are both surprisingly robust to diamagnetic doping; and (iii) an additional contribution at $T$ < 0.4 K appears with increasing $x$. The origin of this unusual behaviour is rationalised in terms of the fragmentation of quasi-1D spin chains by the diamagnetic Y$^{3+}$ dopant. Magnetocaloric measurements show a nonlinear dependence on $x$. The mass-weighted magnetocaloric entropy decreases across the series from the promising values in Tb(HCO$_2$)$_3$; however, the magnetocaloric entropy per magnetic Tb$^{3+}$ ion first decreases then increases with increasing $x$. Our results establish Tb$_{1-x}$Y$_x$(HCO$_2$)$_3$ as a model system in which to explore the functional ramifications of dilution in a low-dimensional magnet.

## 1. Introduction

The magnetocaloric effect (MCE) is an entropically-driven phenomenon that has long been used for ultra-low (sub 2 K) cryogenic cooling. The effect is observed in all magnetic materials upon application and subsequent removal of an external magnetic field:[1] an entropy decrease is caused by aligning magnetic moments in an applied field, and this entropy can be recovered when the field is removed. Magnetocalorics are a promising solid-state alternative to vapour compression refrigeration for cryogenic temperatures. This is particularly the case for the regimes in which liquid helium is conventionally used, due to this being an increasingly scarce and expensive resource.[2] An important development in the field is the recent appreciation that unconventional magnetic order can play a key role in optimising MCE materials for applications at lower applied fields.[3–6]

Historically, Gd-based magnetocalorics have been favoured because Gd$^{3+}$ ($S$ = 7/2) has the largest spin-only maximum magnetic entropy change $\Delta S_m^{max}$ = $R$ln($2S$ + 1) of any magnetic ion.[7] Indeed, a number of Gd-containing oxides and alloys have been reported to show good magnetocaloric properties for low temperature cooling; arguably the most famous is gadolinium gallium garnet (GGG), which is the benchmark material for cooling applications under 10 K.[8] Such materials are, however, often densely-packed structures where strong interactions among the magnetic moments result in the emergence of long-range order at relatively high temperature, commonly restricting their use as magnetocalorics, meaning that they are unsuitable for cooling in lower temperature ranges.

In this context, coordination polymers and metal-organic frameworks (MOFs)—with their versatile structural chemistry—offer an attractive mechanism of optimizing MCE by targeting structures that are predisposed to unconventional magnetic order.[9] For example, it is straightforward to engineer low-dimensional motifs, such as magnetic chains and sheets. When combined with magnetic frustration between these low-dimensional units, the long-range order in these materials can be suppressed to very low temperatures — even in structures with a relatively high density of magnetic ions. Competing magnetic interactions in turn enable the optimisation of magnetocaloric properties when magnetic order is achieved.[9,10] The usual focus is on designing structures featuring triangular motifs, which are responsible for the competition of antiferromagnetic interactions; examples include the triangular, kagome and trillium nets.[6,11,12]

The $Ln$(HCO$_2$)$_3$ coordination frameworks ($Ln$ = Gd$^{3+}$, Tb$^{3+}$, Dy$^{3+}$, Ho$^{3+}$, Er$^{3+}$) comprise one such family of modern MCE materials.[13] These systems have proven to be highly efficient magnetocalorics over a wide temperature range (2-20 K). While Gd(HCO$_2$)$_3$ has the strongest performance at ultra-low temperatures and high applied fields, it has been found that replacing Gd with Ising-like Tb and Ho leads to enhanced performance above 4 K in more modest applied fields of up to 2 T.[13] This reduced field requirement is ideal from an applications perspective since it can be achieved using permanent magnets.[13] The attractive properties of Tb(HCO$_2$)$_3$ can be attributed to their highly anisotropic and frustrated magnetism: strongly-coupled ferromagnetic ($J_{intrachain}$ = 1.5(5) K) Tb$^{3+}$ chains are packed on a triangular lattice with much weaker and antiferromagnetic inter-chain coupling ($J_{interchain}$ = 0.03(1) K).[14] It is the frustration of this state that allows chain magnetisation to be reversed easily under low applied fields.

Tb(HCO$_2$)$_3$ enters a partially-ordered triangular Ising antiferromagnet (TIA) state below 1.6 K,[14,15] in which ferromagnetic long-range order is achieved along the direction of the chains, with short-range order between them due to the frustrated triangular lattice. Consistent with the proposed origin of the magnetocaloric properties of Tb(HCO$_2$)$_3$, neutron diffraction studies have shown that the application of weak magnetic fields helps to stabilise the TIA phase; however, at fields above 0.2 T, the system transforms to a simple ferromagnetic state in which all chains are aligned.[16]

It is well-known that diamagnetic impurities readily disrupt low-dimensional magnetic states,[17–21] so the unusual discovery of the TIA state in Tb(HCO$_2$)$_3$ raises the question of how this state responds to the introduction of such impurities, and in turn how diamagnetic impurities affect the magnetocaloric properties observed. In principle, replacing Tb$^{3+}$ with diamagnetic Y$^{3+}$ should be viable, given that the two cations have similar ionic radii: 1.095 and 1.075 Å for 9-coordinated Tb$^{3+}$ and Y$^{3+}$, respectively.[22] Indeed we report here the synthesis and characterization of a new family of Y-doped MOFs, the Tb$_{1-x}$Y$_x$(HCO$_2$)$_3$ frameworks. Heat capacity measurements show that the TIA state in this family is surprisingly robust to the introduction of diamagnetic impurities, and we discover an unexpected additional contribution to the low-temperature heat capacity at higher Y compositions. We ascribe this additional entropic contribution to the effect of spin-chain fragmentation. This may be related to the unusual experimental observation that, while a decrease in magnetocaloric properties is observed initially with Y doping, higher levels of diamagnetic defect density actually enhance the per-spin magnetocaloric effect. We discuss the implications of our results in the context of MCE materials.

2. **Experimental**

Samples of Tb$_{1-x}$Y$_x$(HCO$_2$)$_3$ ($x$ = 0.025, 0.05, 0.10, 0.20, 0.40 0.60, 0.80) were synthesised following a previously reported procedure.[15] Combinations of Y(NO$_3$)$_3$·6H$_2$O (99.9%, Acros Organics) and Tb(NO$_3$)$_3$·6H$_2$O (99%, Acros Organics) in different molar ratios were dissolved in 4.75 ml formic acid (99%, Acros Organics) with the addition of 0.25 ml ethanol used to slow down the reactions. After several minutes of stirring, $NO_y$ was released, and white products precipitated out of solution. The products were filtered under vacuum and washed several times using ethanol.

Sample purity was assessed by powder X-ray diffraction (PXRD) using a Bragg-Brentano PANalytical X'PERT 3 diffractometer equipped with an Empyrean CuK$_α$ LFF source (λ = 1.5046 Å) and a X'Celerator linear detector with the samples mounted on zero-background silicon sample holders. The resulting patterns were analysed for phase purity using the program Rietica, employing the Le Bail fitting method.[23,24]

Compositional analysis of the cations present in the $Tb_{1-x}Y_x(HCO_2)_3$ series was carried out via energy dispersive X-Ray Fluorescence (EDXRF) measurements using a PANalytical Epsilon 3 spectrometer. Results obtained were calibrated by using a calibration curve determined from physically ground mixtures of $Y(HCO_2)_3$ and $Tb(HCO_2)_3$ with different Tb:Y molar ratios; namely, 97.5:2.5, 95:5, 90:10, 75:25, 50:50 and 27:75. Scanning Electron Microscopy (SEM) measurements and further energy-dispersive X-ray spectroscopy analysis (EDX) on selected regions of the samples were carried out using a Hitachi S3400N microscope. Powder samples were deposited on carbon tabs placed on aluminium sample holders, without further preparation, such as coating or surface polishing. Images of regions of approximately 56 x 44 μm$^2$ were acquired using a 10 kV electron beam and a secondary electron (SE) detector.

DC susceptibility and isothermal magnetisation measurements for $Tb_{1-x}Y_x(HCO_2)_3$, with x = 0.025, 0.05, 0.10, were performed using a Quantum Design MPMS or MPMS-3 SQUID Magnetometer, with powder samples placed in gelatine capsules placed within pierced straws with a diamagnetic background. Heat capacity data between 4 and 14 K for $Tb(HCO_2)_3$ and 250 mK and 4 K for $Tb(HCO_2)_3$ and $Tb_{1-x}Y_x(HCO_2)_3$, with $x$ = 0.05, 0.10, 0.20 and 0.40 were collected using a Quantum Design PPMS-Dynacool at the ISIS support Laboratories, Rutherford Appleton Laboratories, UK.

## 3. Results and Discussion

In order to determine the compositions of our $Tb_{1-x}Y_x(HCO_2)_3$ samples, we used X-ray fluorescence (XRF) measurements. Our results suggest the experimental compositions are close to the nominal ones expected from the synthesis (see Table 1). EDX measurements were also performed on selected regions of each sample, confirming there was no indication of significant inhomogeneity in the distribution of metals in the materials (see Fig. S1-S7 and Tables S1-S7). There was some small variation between different regions of the sample, which can be primarily attributed to the surface roughness of the powder samples and which is accounted for in the uncertainty of the measurements. Overall, the average stoichiometries are in good agreement with those determined from bulk XRF analysis. Nominal stoichiometries have been used for the normalisation of our data per amount of Tb present in each sample.

Our room-temperature powder X-ray diffraction measurements show that all the members of the series share the same rhombohedral $R$3$m$ structure as the parent $Tb(HCO_2)_3$ and $Y(HCO_2)_3$ phases (see Fig. 1). Le Bail fits to powder X-ray diffraction patterns confirmed phase purity of our samples, with the absence of peak broadening upon higher Y doping suggesting relatively high sample homogeneity for all the samples (see Table S8 for refinement statistics). In this structure, $Ln$O$_9$ coordination polyhedra form face-sharing chains propagating down the $c$-axis. Neighbouring chains are then joined to one another $via$ formate ligands, resulting in a triangular arrangement within the $ab$-plane. [13–15] The lattice parameters decrease with higher Y doping, as would be expected given the smaller size of $Y^{3+}$, showing subtle deviations from Vegard's law which may indicate some degree of dopant anticlustering (see Fig. S8).[25]

**Table 1.** Comparison of the nominal and experimental compositions for $Tb_{1-x}Y_x(HCO_2)_3$ determined from X-ray fluorescence.

| Nominal composition | Experimental composition |
| --- | --- |
| $Tb_{0.975}Y_{0.025}(HCO_2)_3$ | $Tb_{0.99(6)}Y_{0.01(6)}(HCO_2)_3$ |
| $Tb_{0.95}Y_{0.05}(HCO_2)_3$ | $Tb_{0.96(5)}Y_{0.04(5)}(HCO_2)_3$ |
| $Tb_{0.90}Y_{0.10}(HCO_2)_3$ | $Tb_{0.91(4)}Y_{0.09(4)}(HCO_2)_3$ |
| $Tb_{0.80}Y_{0.20}(HCO_2)_3$ | $Tb_{0.819(19)}Y_{0.181(19)}(HCO_2)_3$ |
| $Tb_{0.60}Y_{0.40}(HCO_2)_3$ | $Tb_{0.63(3)}Y_{0.37(3)}(HCO_2)_3$ |
| $Tb_{0.40}Y_{0.60}(HCO_2)_3$ | $Tb_{0.44(8)}Y_{0.56(8)}(HCO_2)_3$ |
| $Tb_{0.20}Y_{0.80}(HCO_2)_3$ | $Tb_{0.25(12)}Y_{0.75(12)}(HCO_2)_3$ |

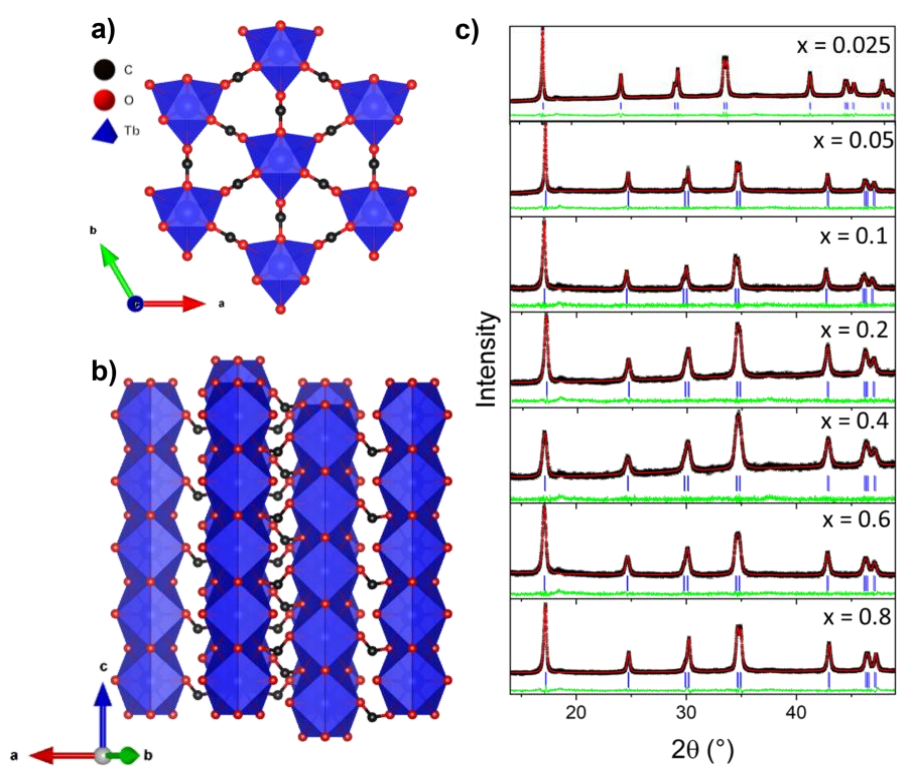

**Fig. 1:** Crystal structure of $Tb(HCO_2)_3$ with a) one-dimensional infinite chains connected by formate ligands and arranged on a triangular lattice on the *ab*-plane and b) face-sharing chains propagating along the *c*-axis. c) Conventional powder X-ray diffraction patterns ($\lambda = 1.5046$ Å) of $Tb_{1-x}Y_x(HCO_2)_3$ fitted using the Le Bail method to highlight phase purity. The crosses, red and green lines are experimental and calculated intensities and the difference curve. Vertical markers indicate the positions of the Bragg reflections.

Field cooled (FC) magnetic susceptibility $\chi(T)$ measurements of the $Tb_{1-x}Y_x(HCO_2)_3$ frameworks, with *x* = 0.025, 0.05, 0.10, were carried out in a 0.1 T field from 2 K to 300 K (see Fig. 2a and Fig. S9-S11). These susceptibility data did not show any indication of long-range magnetic ordering within this temperature range. This observation is consistent with previous magnetic susceptibility measurements on $Tb(HCO_2)_3$, for which no deviation from paramagnetic behaviour is observed to well below 2 K.[13–15] The inverse susceptibilities of our $Tb_{1-x}Y_x(HCO_2)_3$ samples were well fitted using the Curie-Weiss law from 10 to 300 K; the corresponding $\theta_{CW}$ values are given in Table 2. From mean-field theory, one expects $\theta_{CW}$ to scale with $Tb^{3+}$ concentration if there is no variation in the strength of magnetic interactions on dilution by diamagnetic ions. Our data reflect this trend, and so we assume that the coupling among $Tb^{3+}$ spins is likely to be similar across the whole series to that present in the parent Tb(HCO$_2$)$_3$ framework — i.e. ferromagnetic intrachain and antiferromagnetic interchain coupling. Effective magnetic moments $\mu_{eff}$ were found to be close to the reported experimental value for Tb(HCO$_2$)$_3$ and also the 9.75 $\mu_B$ value expected from the Russel-Saunders coupling scheme (see Table 2).[13]

Magnetisation measurements from 12 K down to 2 K are consistent with paramagnetic behaviour (see Fig. S12-S16). Observed normalised values of saturation magnetisation $M_{sat}$ at 2 K and a magnetic field of 5 T are close to the theoretical value of 4.50 $\mu_B$ atom$_{Tb}^{-1}$ expected for Ising behaviour, for which $M_{sat}$ is expected to be close to $g_JJ/2$, suggesting significant single ion anisotropy. They are, however, significantly lower than the reported value of 5.77(4) $\mu_B$ atom$^{-1}$ previously reported for Tb(HCO$_2$)$_3$ (see Fig. 2b and Table 2).[13]

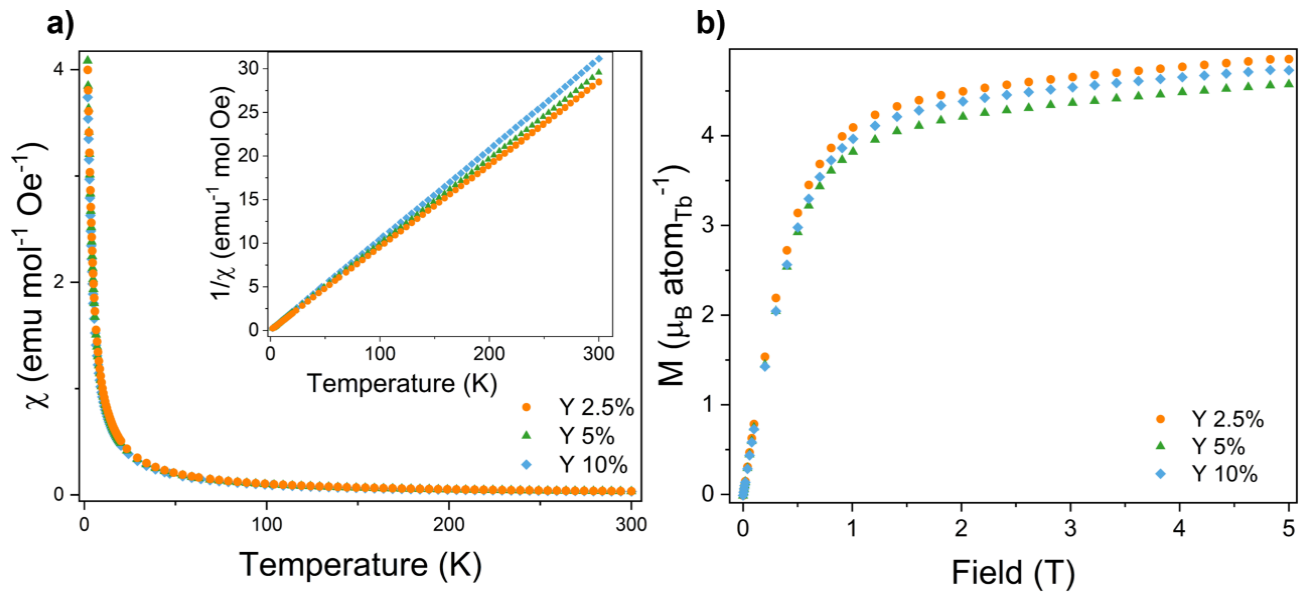

**Fig. 2:** a) FC molar susceptibility $\chi(T)$ for Tb$_{1-x}$Y$_x$(HCO$_2$)$_3$ with $x$ = 0.025, 0.05, 0.10, measured from 2 – 300 K in a field of 0.1 T; the inverse molar susceptibility $\chi^{-1}(T)$ is presented in the inset. b) Magnetisation plot of Tb$_{1-x}$Y$_x$(HCO$_2$)$_3$ ($x$ = 0.025, 0.05 and 0.10) collected at 2 K for fields up to 5 T.

**Table 2.** Magnetic properties extracted from FC susceptibility and magnetisation measurements of Tb$_{1-x}$Y$_x$(HCO$_2$)$_3$ samples. Magnetic properties for $x$ = 0 were measured from Saines *et al.*[13]

| $x$ | $\theta_{CW}$ (K) | $|\theta_{CW}|/(1-x)$ (K) | Tb$^{3+}$ Magnetic moment ($\mu_B$ atom$_{Tb}^{-1}$) | $M_{sat}$ ($\mu_B$ atom$^{-1}$) | $M_{sat}$ ($\mu_B$ atom$_{Tb}^{-1}$) |
|---|---|---|---|---|---|
| 0.0 | −0.9 | 0.9 | 9.62 | 5.77(4) | 5.77(4) |
| 0.025 | −0.86(5) | 0.88(5) | 9.417(2) | 4.73(2) | 4.86(2) |
| 0.05 | −1.11(8) | 1.17(8) | 9.511(2) | 4.34(2) | 4.57(2) |
| 0.10 | −0.79(8) | 0.88(8) | 9.771(2) | 4.25(2) | 4.73(2) |

Heat capacity $C(T)$ data were measured for Tb(HCO$_2$)$_3$ and Tb$_{1-x}$Y$_x$(HCO$_2$)$_3$ with $x$ = 0.05, 0.10, 0.20 and 0.40 in zero-field conditions. The heat capacity includes nuclear, electronic, lattice and magnetic contributions: $C(T) = C_{nuc} + C_{el} + C_{lat} + C_{mag}$. Here, the electronic term $C_{el}$ can safely be ignored as the materials are insulators. The lattice contribution $C_{lat}$ was estimated using a Debye fit to the heat capacity data for Tb(HCO$_2$)$_3$ between 8 to 14 K, where phonon population is the dominant process at play. We obtained a Debye temperature of 156.2(1.8) K, which we used for all values of $x$.[26] The value of $C_{lat}$ is very small at the temperatures of relevance to magnetic ordering ($T$ < 4 K) (see Fig. S17). Finally, it was possible to estimate the nuclear contribution $C_{nuc}$ due to the Schottky anomaly, which arises below 0.8 K as a result of magnetic hyperfine coupling of the electronic spins in the 4$f$ orbitals of terbium and the magnetic moment of the terbium nucleus ($I$ = 3/2). This contribution was estimated by fitting to the low-temperature data 0.4 < $T$ < 0.8 K using the expression:

$$C_{\text{nuc}} = \frac{\Delta^2 R}{9T^2} \frac{\exp(-\Delta/3T)[1+\exp(-\Delta/3T)]^4 + 4\exp(-\Delta/T)}{[1+\exp(-\Delta/3T)+\exp(-2\Delta/3T)+\exp(-\Delta/T)]^2}$$

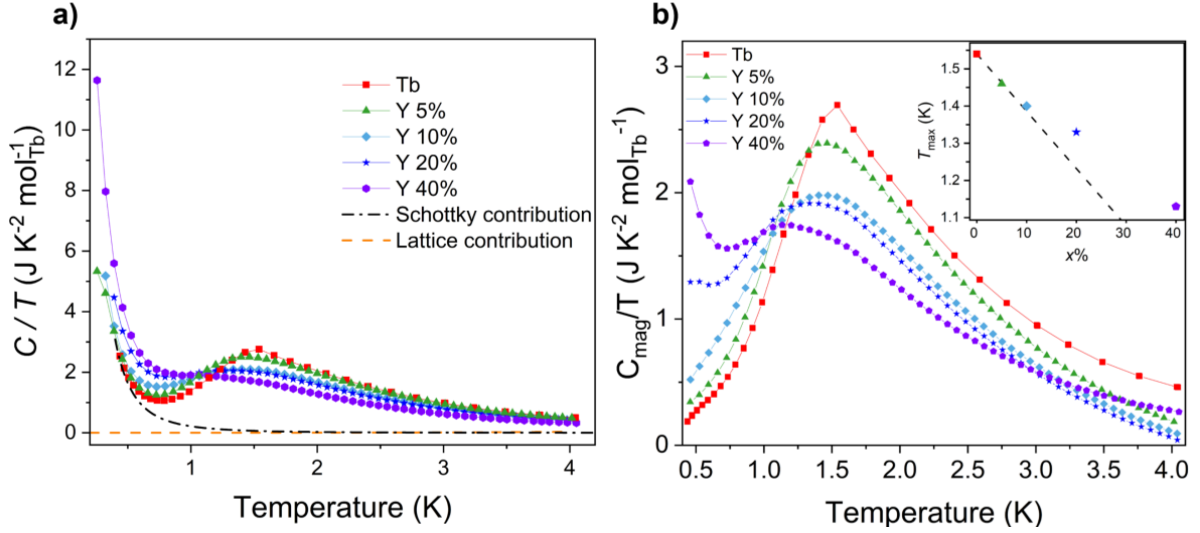

**Fig. 3:** Plot of magnetic heat capacity $C_{\text{mag}}/T$, normalised per moles of Tb, for Tb$_{1-x}$Y$_x$(HCO$_2$)$_3$ ($x$ = 0, 0.05, 0.10, 0.20, 0.40) (a) measured between 300 mK and 4 K in zero field conditions and (b) after subtracting the hyperfine coupling contribution for the Tb cations. In the latter case, the data is shown down to 0.4 K, the lowest limit for the fits for the hyperfine coupling. Values of $T_{\text{max}}$ are shown in the inset, with the dashed line representing the trend as predicted by mean field theory.

where the single adjustable parameter Δ captures the hyperfine splitting.[27] We found Δ = 0.437(2) K for Tb(HCO$_2$)$_3$, which is consistent with many other Tb-containing systems, including Tb metal itself (see Table S9)[27–29] suggesting this value solely depends on the internal electronic structure of Tb and is independent of the local chemical environment. Therefore, this parameter is here assumed to be constant with respect to stoichiometry and subtracted from the measured heat capacity for all samples. Doing so leads to the curious result that this hyperfine coupling contribution does not account for the entirety of the observed low-temperature heat capacity of those members with higher concentrations of Y$^{3+}$ (see Fig. 3). The resulting $C_{\text{mag}}(T)$ data, obtained following subtraction of lattice and nuclear contributions, show that Tb(HCO$_2$)$_3$ has the strongest magnetic contribution to the heat capacity, with this contribution decreasing with higher Y$^{3+}$ concentration in the other samples, as would be expected.

The values for the temperatures of the maxima of the heat capacity curves, indicative of the transition temperature and the strength of the correlations within the system, were extracted and these were found to be centred at 1.54 K for Tb(HCO$_2$)$_3$, close to the reported transition temperature to the TIA state for this system (see inset Fig 3b).[15] There is a surprisingly modest decrease in this transition temperature with Y$^{3+}$ doping. Even where almost half of the Tb$^{3+}$ cations are replaced with Y$^{3+}$ in Tb$_{0.60}$Y$_{0.40}$(HCO$_2$)$_3$, the TIA transition still occurs at a similar temperature, being reduced by less than 30 % of the undoped Tb(HCO$_2$)$_3$. Further heat capacity measurements were performed under different applied fields on Tb$_{1-x}$Y$_x$(HCO$_2$)$_3$ members with x = 0.10, 0.20 and 0.40. These data show that the magnetic component of the heat capacity $C_{\text{mag}}$ is suppressed with higher applied fields, confirming that the features observed are magnetic in nature (see Fig. S18-S20).

The low-temperature magnetic entropy loss $S_{\text{rel}}$, was extracted from the zero-field heat capacity data using the thermodynamic relationship $S_{\text{rel}}(T) = -\int_T^{T_{max}} C_{\text{mag}}(T')/T' \, dT'$, with $T_{\text{max}}$ = 4 K chosen as useful limiting temperature well above the TIA transition (see Fig. 4 and Table 3). The total loss down to 0.5 K (the lowest temperature point for which we have reliable data), which we call the maximal magnetic entropy $S_{rel}^{max}$, decreases with magnetic dilution more quickly than the TIA ordering temperature — even when normalised by the amount of Tb present. The shape of the $S_{rel}(T)$ function varies as $x$ is increased, which is a result of the low-temperature specific-heat feature observed for Y-doped samples.

Our key observations from heat capacity measurements of the $Tb_{1-x}Y_x(HCO_2)_3$ series are (i) that the peak at ~1.5 K, associated with formation of the TIA state in $Tb(HCO_2)_3$, broadens and shifts to only moderately lower temperature with $Y^{3+}$ doping, and (ii) that there is an additional low-temperature contribution to the heat capacity at high levels of diamagnetic doping fraction. The former indicates that the formation of the TIA state caused by the ordering of the 1D chains within $Tb_{1-x}Y_x(HCO_2)_3$ is surprisingly robust to magnetic dilution. The entropy loss associated with formation of the TIA state — i.e. the area under the main broad peak in $C_{mag}/T$ — decreases in magnitude with increasing $x$; this affect is stronger than the reduction in $T_{max}$ and suggests the extent of magnetic ordering is affected more by doping than the magnetic transition temperature. We interpret (ii) as arising from an entropic contribution due to the presence of short chains, which are a result of the $Y^{3+}$ ions breaking the strongly interacting quasi-1D chains into small segments. Because of the much weaker interchain interactions, we would not expect these short chains to order until significantly below the TIA transition, at which temperature they contribute to the upturn in $C_{mag}/T$ seen at the lowest temperatures. Simple statistical arguments, assuming no dopant clustering, give an entropy of $S_{rel}(x) = -(1-x)R\ln 2$ associated with complete intra-short-chain order. We find good qualitative agreement to the experimental values of $S_{rel}(T)$ evaluated at 0.7 K, the relevant turning point in $C_{mag}/T$ (see inset to Fig. 4). Since our integral only spans a portion of this in-chain ordering, $S_{rel}(x) = -(1-x)S_{rel}(0)$ is a more appropriate comparison. We intend to investigate the behaviour of these short chains in more detail in a forthcoming computational study.

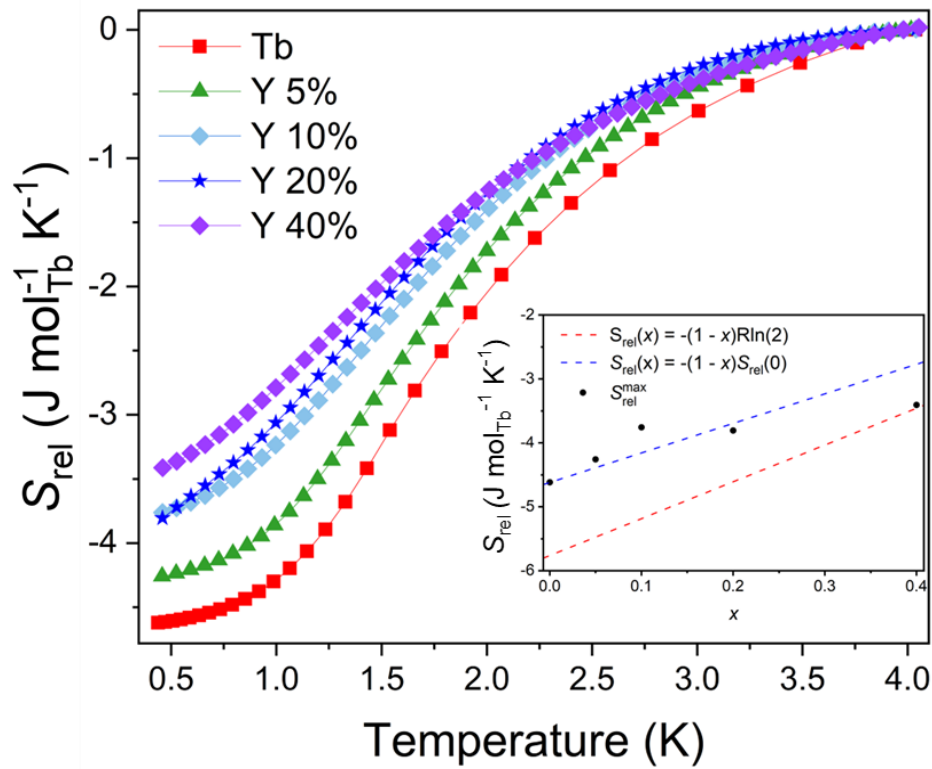

**Fig. 4:** Magnetic entropies per moles of Tb extracted from the heat capacity data for the $Tb_{1-x}Y_x(HCO_2)_3$ ($x$ = 0, 0.05, 0.10, 0.20, 0.40) series in zero-field conditions. The inset shows the maximal values and the $S_{rel}$ models derived from statistics.

**Table 3.** Maximal magnetic entropy loss associated with the TIA transition for the Tb$_{1-x}$Y$_x$(HCO$_2$)$_3$ solid solutions with $x = 0$, 0.05, 0.10, 0.20 and 0.40.

| Sample | -S$_{rel}^{max}$ (J kg$^{-1}$ K$^{-1}$) | -S$_{rel}^{max}$ (J mol$_{Tb}^{-1}$ K$^{-1}$) |
|---|---|---|
| Tb(HCO$_2$)$_3$ | 15.72 | 4.62 |
| Tb$_{0.95}$Y$_{0.05}$(HCO$_2$)$_3$ | 13.93 | 4.26 |
| Tb$_{0.90}$Y$_{0.10}$(HCO$_2$)$_3$ | 11.79 | 3.76 |
| Tb$_{0.80}$Y$_{0.20}$(HCO$_2$)$_3$ | 10.86 | 3.81 |
| Tb$_{0.60}$Y$_{0.40}$(HCO$_2$)$_3$ | 7.69 | 3.41 |

The magnetic entropy change -$\Delta S_m$ as a function of applied magnetic field was extracted from the magnetisation data using the Maxwell relation $\Delta S_m(T) = \int [\delta M(T, H)/\delta T]_H \, dH$ from 2 K to 12 K for $x$ = 0.025, 0.05, 0.10, 0.20 and 0.40 (see Fig. 5, Fig. S20 and Fig. S21 for $\Delta B$ = 5-0 and 2-0 T field changes; see Fig. S23 and Fig. S24 for $\Delta B$ = 1-0 T. See Table 4 for maximum entropy changes). In determining the volumetric values, the crystallographic densities of each of the measured members of the series were estimated by calculating the ratio between the formula weights of each member and the unit cell volumes determined from the Le Bail fits to the powder diffraction data (see Table S10 for density used).[22] Compared to the values reported for Tb(HCO$_2$)$_3$,[13] those found for the doped samples clearly show that the magnetocaloric performance of this material decreases due to Y-doping as a function of sample weight or volume. However, the gravimetric entropy changes of the $x$ = 0.40 sample increase again for 2-0 and 1-0 T field changes compared to the x = 0.20 compound. Interestingly, normalised values per mole of Tb$^{3+}$ are lower compared to the -$\Delta S_m^{max}$ for the parent compound Tb(HCO$_2$)$_3$, with the decrease in -$\Delta S_m^{max}$ with respect to Y-doping more dramatic for low values of $x$ — particularly for more modest applied field changes. This trend is then reversed at higher concentrations, with Tb$_{0.60}$Y$_{0.40}$(HCO$_2$)$_3$ having the highest -$\Delta S_m^{max}$ amongst the Y doped samples with respect to the amount of Tb$^{3+}$ present for all field changes considered. Exceptionally Tb$_{0.60}$Y$_{0.40}$(HCO$_2$)$_3$ has a higher normalised -$\Delta S_m^{max}$ than that of the parent compound Tb(HCO$_2$)$_3$ for all the applied field changes (see insets in Fig.5 and Fig S.23, as well as Fig. S21, S22 and S24).

When compared to other magnetically-dilute MCE materials, such as the Ga-doped La$_{0.7}$(Ba, Sr)$_{0.3}$Mn$_{1-x}$Ga$_x$O$_3$ ($x$ = 0, 0.1, 0.2),[30] the Cr-doped La$_{0.65}$Eu$_{0.05}$Sr$_{0.3}$Mn$_{1-x}$Cr$_x$O$_3$ ($x$ = 0.05, 0.1, and 0.15),[31] and the Sn-substituted RCo$_2$ (R = Gd, Tb, Dy) alloys, the Tb$_{1-x}$Y$_x$(HCO$_2$)$_3$ series follows similar trends with respect to its initial decrease in -$\Delta S_m^{max}$ with Y$^{3+}$ doping — i.e. lower transition temperatures and lower values of $-\Delta S_m^{max}$ when higher concentrations of non-magnetic impurities are present. This has also been observed in the geometrically frustrated Tm$_{1-x}$Lu$_x$B$_4$, which features a Shastry-Sutherland lattice, for $x$ < 0.06.[32] However, for Tm$_{0.7}$Lu$_{0.3}$B$_4$ the trend is reversed. Monte-Carlo modelling suggested that in Tm$_{1-x}$Lu$_x$B$_4$ small concentrations of non-magnetic ions relieve the degeneracy of the ground state for this magnetic system, which results in a decrease in ground-state entropy and, therefore, of the MCE. Conversely, higher concentrations of non-magnetic dopant result in the formation of nearly-independent clusters which are responsible for a significant degeneracy of the ground-state, leading to a higher zero-field entropy change and, ultimately, in an improvement of the MCE.[32] This observation is compatible with the putative role of fragmented chains discussed here.[32] Indeed, theoretical studies of selective doping of the sublattices of a Triangular Ising Antiferromagnet have shown that it is possible to achieve a similar effect, i.e. an increase in the maxima of entropy change when higher concentration of non-magnetic dopants are present in such a system.[33]

These results should be understood in the context of the promising magnetocaloric behaviour of Tb(HCO$_2$)$_3$ emerging from the frustration of its strongly coupled ferromagnetic chains leading to a significant zero-field entropy, which is suppressed on application of relatively small fields. We propose that the initial decrease in normalised -$\Delta S_m^{max}(x)$ with increasing x is caused by the rapid suppression of this geometric frustration due to local symmetry breaking. This is reflected in the initial decrease in normalised -$\Delta S_m^{max}$ being somewhat greater for more modest applied fields. Conversely, as the concentration of $Y^{3+}$ increases further there are likely a higher number of short chains present, which leads to a significantly more disordered state in the absence of an applied magnetic field and thus a greater change in the normalised -$\Delta S_m^{max}$.

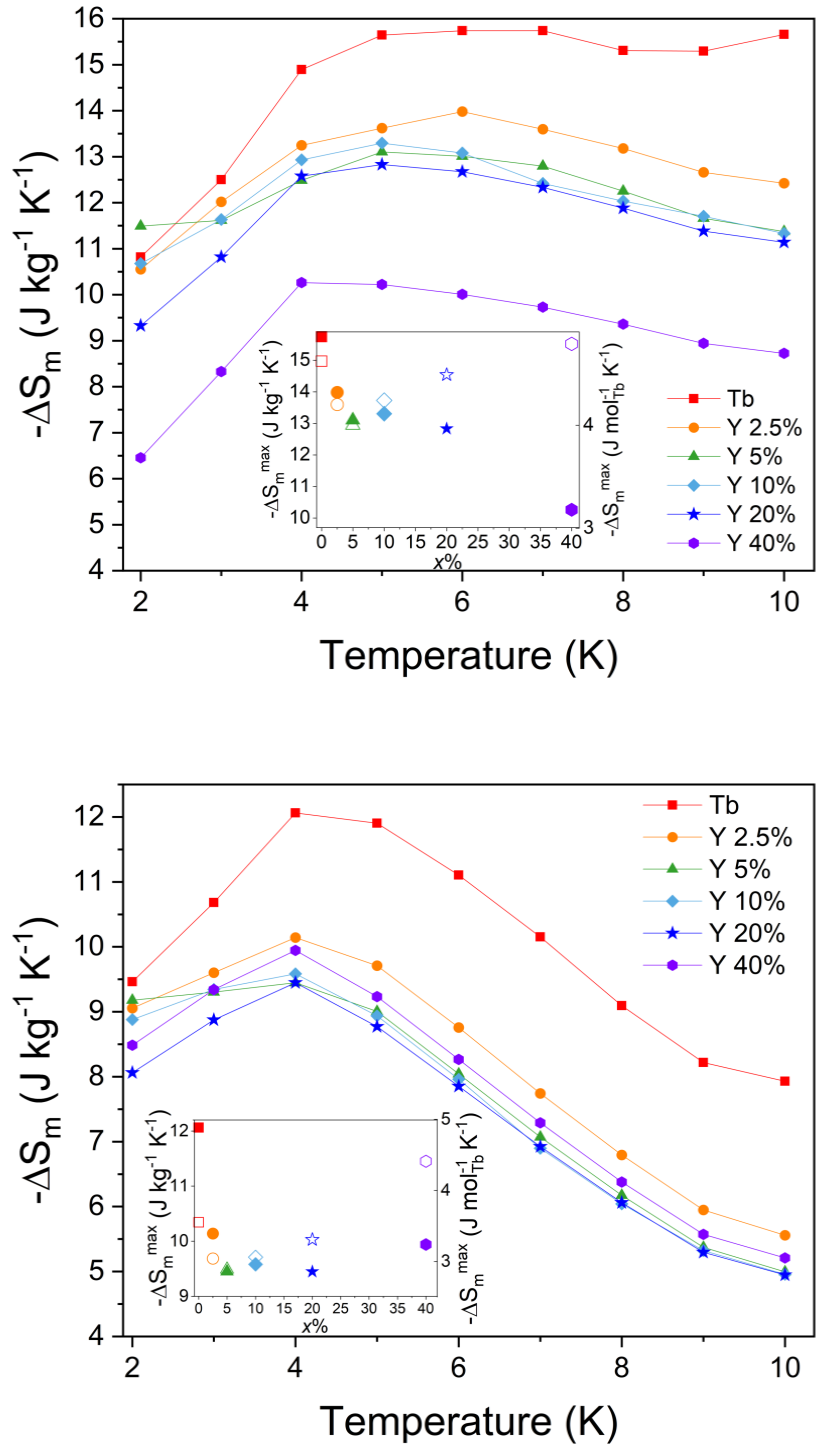

**Fig. 5:** Gravimetric magnetic entropy changes extracted from the magnetisation data for $Tb_{1-x}Y_x(HCO_2)_3$ ($x$ = 0, 0.025, 0.05, 0.10, 0.20, 0.40) series for $\Delta B$ = 5-0 T (top) and $\Delta B$ = 2-0 T (bottom). Insets show the values of -$\Delta S_m^{max}$ in gravimetric (full symbols) and normalised molar (hollow symbols) units for each sample.

**Table 4.** Maximum entropy changes extracted from magnetisation data for the $Tb_{1-x}Y_x(HCO_2)_3$ solid solutions ($x$ = 0, 0.025, 0.05, 0.10) for field changes of 5-0, 2-0 and 1-0 T.

| Sample | $\Delta B$ = 5-0 T | | | $\Delta B$ = 2-0 T | | | $\Delta B$ = 1-0 T | | |
|---|---|---|---|---|---|---|---|---|---|
| | $-\Delta S_m^{max}$ (J kg$^{-1}$ K$^{-1}$) | $-\Delta S_m^{max}$ (mJ cm$^{-3}$ K$^{-1}$) | $-\Delta S_m^{max}$ (J mol$_{Tb}^{-1}$ K$^{-1}$) | $-\Delta S_m^{max}$ (J kg$^{-1}$ K$^{-1}$) | $-\Delta S_m^{max}$ (mJ cm$^{-3}$ K$^{-1}$) | $-\Delta S_m^{max}$ (J mol$_{Tb}^{-1}$ K$^{-1}$) | $-\Delta S_m^{max}$ (J kg$^{-1}$ K$^{-1}$) | $-\Delta S_m^{max}$ (mJ cm$^{-3}$ K$^{-1}$) | $-\Delta S_m^{max}$ (J mol$_{Tb}^{-1}$ K$^{-1}$) |
| $Tb(HCO_2)_3$ | 15.74 | 61.62 | 4.62 | 12.06 | 47.23 | 3.55 | 8.08 | 31.64 | 2.38 |
| $Tb_{0.975}Y_{0.025}(HCO_2)_3$ | 13.98 | 54.04 | 4.20 | 10.14 | 39.21 | 3.04 | 6.42 | 24.82 | 1.92 |
| $Tb_{0.95}Y_{0.05}(HCO_2)_3$ | 13.10 | 50.39 | 4.00 | 9.45 | 36.33 | 2.89 | 6.15 | 23.64 | 1.88 |
| $Tb_{0.90}Y_{0.10}(HCO_2)_3$ | 13.29 | 50.67 | 4.24 | 9.58 | 36.52 | 3.06 | 6.07 | 23.13 | 1.93 |
| $Tb_{0.80}Y_{0.20}(HCO_2)_3$ | 12.83 | 47.85 | 4.49 | 9.45 | 35.24 | 3.31 | 5.77 | 21.52 | 2.02 |
| $Tb_{0.60}Y_{0.40}(HCO_2)_3$ | 10.80 | 38.34 | 4.79 | 9.95 | 35.31 | 4.41 | 6.07 | 21.56 | 2.69 |

## Conclusion

In this work we have reported the synthesis of members of the $Tb_{1-x}Y_x(HCO_2)_3$ series of coordination frameworks. We have monitored the evolution of the magnetic and magnetocaloric properties of these materials upon changing the composition of the sample *via* both magnetisation and heat capacity measurements. The latter have allowed us to observe the low temperature evolution of the TIA transition with the addition of diamagnetic impurities. Maxima in the heat capacity are observed in data measured for all samples up to $x$ = 0.40, confirming these undergo a transition to the TIA state, with the transition shifting to slightly lower temperatures. These results indicate the emergence of the TIA state is quite robust and only slightly affected by the presence of low concentrations of diamagnetic impurities, although the extent of 1D order in this state likely decreases with $Y^{3+}$ doping as indicated by the smaller magnetic entropies. Furthermore, fits to the data have shown an additional contribution to the magnetic heat capacity is present, with this increasing with the amount of Y in the sample. This contribution is ascribed to an entropic effect caused by the increasing presence of shorter segments along the chains of the $Tb_{1-x}Y_x(HCO_2)_3$.

We have found the incorporation of diamagnetic ions to reduce the magnetocaloric properties of these compounds, determined for low Y-doped samples, namely $x$ = 0.025, 0.05 0.10, 0.2 and 0.4, with values of entropy change $-\Delta S_m^{max}$ reduced in magnitude relative to the parent material, $Tb(HCO_2)_3$ — particularly for lower applied fields. However, amongst doped samples, the lowest values of $-\Delta S_m^{max}$, normalised per amount of Tb present, are obtained for $x$ = 0.05, while the values of entropy change increase significantly for higher values of x, independent of the applied fields. We have proposed a mechanism for this effect, resulting from two competing implications of diamagnetic doping, but thorough theoretical investigation of this effect is required. Overall, this work has highlighted that while the emergence of the TIA state in $Tb(HCO_2)_3$ is surprisingly robust to the dilution of magnetic cations responsible for it by diamagnetic impurities the disruption of the 1D ferromagnetic chains that play a key role in its emergence leads to an unusual trend in the magnetocaloric properties when normalised for the amount of magnetic cation present. As mentioned, we plan to gain further insight about the behaviour of these systems *via* computational studies.

# Electronic Supplementary Information

# Anomalous evolution of the magnetocaloric effect in dilute triangular Ising antiferromagnets $Tb_{1-x}Y_x(HCO_2)_3$

Mario Falsaperna[1], Johnathan M. Bulled[2], Gavin G.B. Stenning[3], Andrew L. Goodwin[2], Paul J. Saines[1].

[1]School of Chemistry and Forensic Science, Ingram Building, University of Kent, Canterbury, CT2 7NH, UK

[2]Department of Chemistry, Inorganic Chemistry Laboratory, University of Oxford, South Parks Road, Oxford OX1 3QR, UK

[3]ISIS Neutron and Muon Source, STFC Rutherford Appleton Laboratory, Chilton, Didcot, OX11 0QX

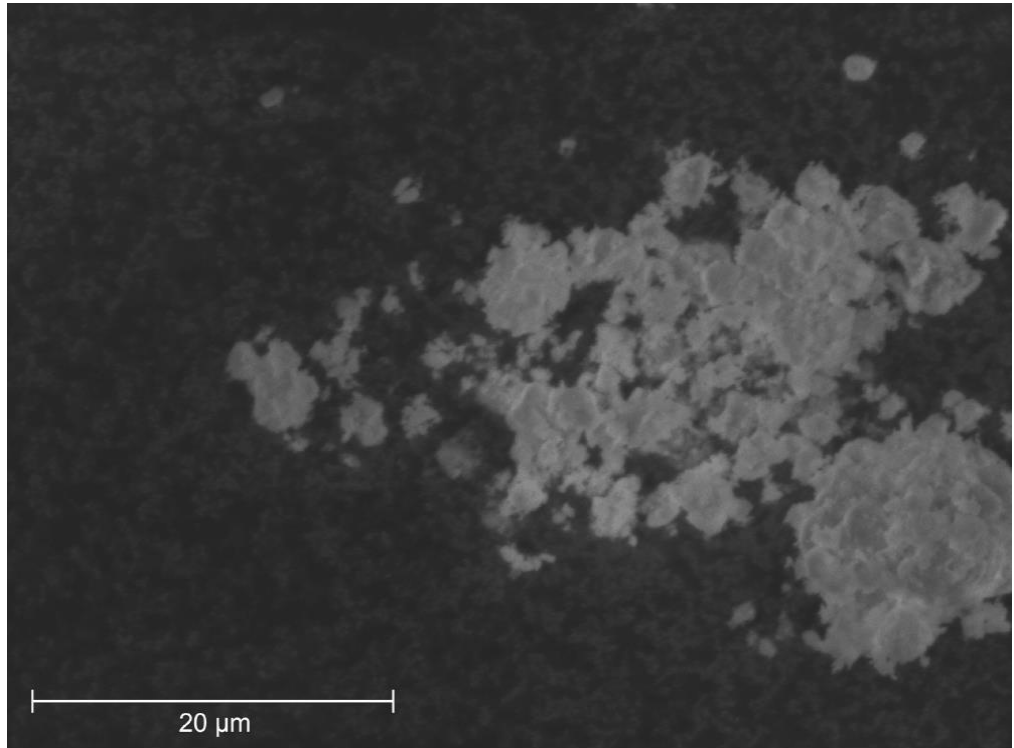

**Fig. S1:** SEM image of $Y_{0.025}Tb_{0.975}(HCO_2)_3$ acquired with a secondary electron (SE) detector.

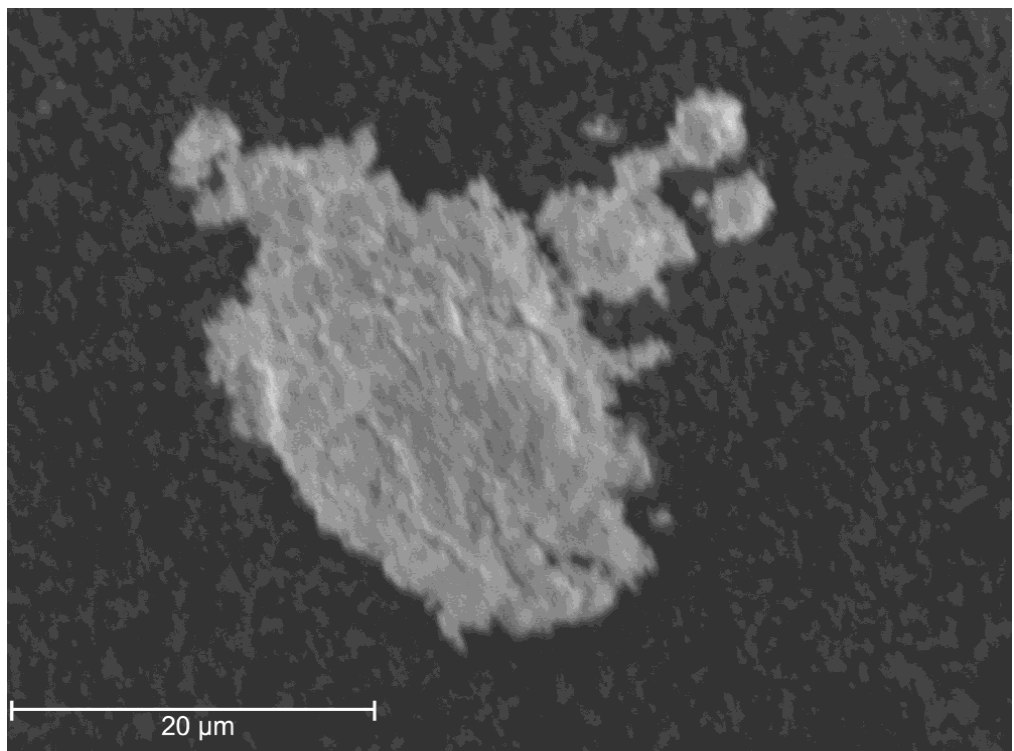

**Fig. S2**: SEM image of $Y_{0.05}Tb_{0.95}(HCO_2)_3$ acquired with a secondary electron (SE) detector.

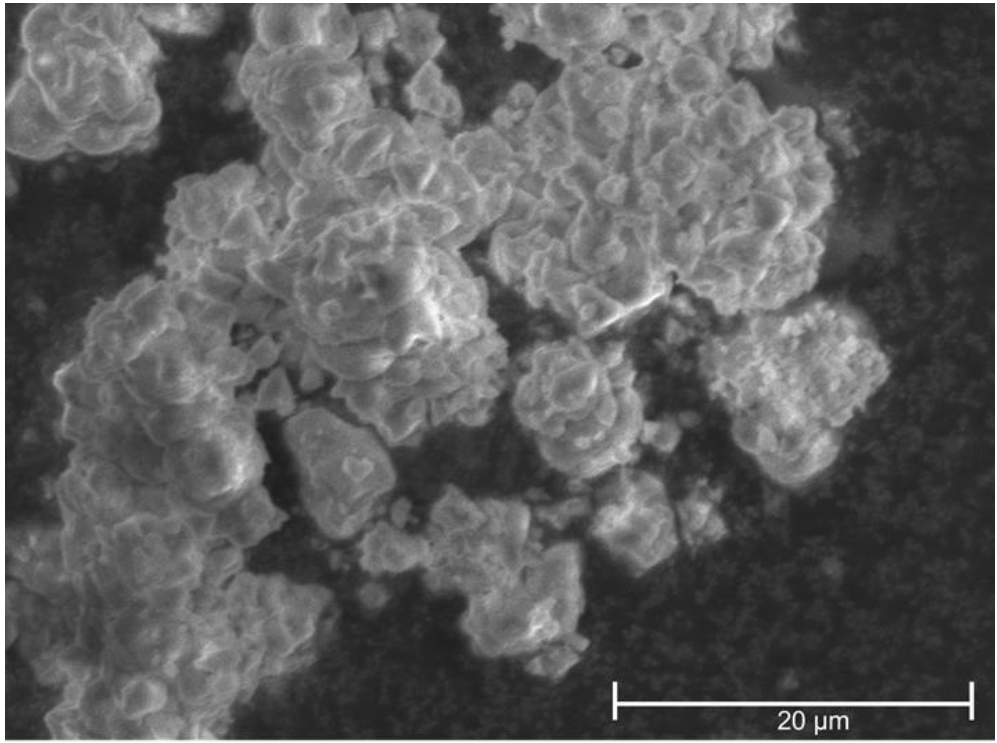

**Fig. S3**: SEM image of $Y_{0.10}Tb_{0.90}(HCO_2)_3$ acquired with a secondary electron (SE) detector.

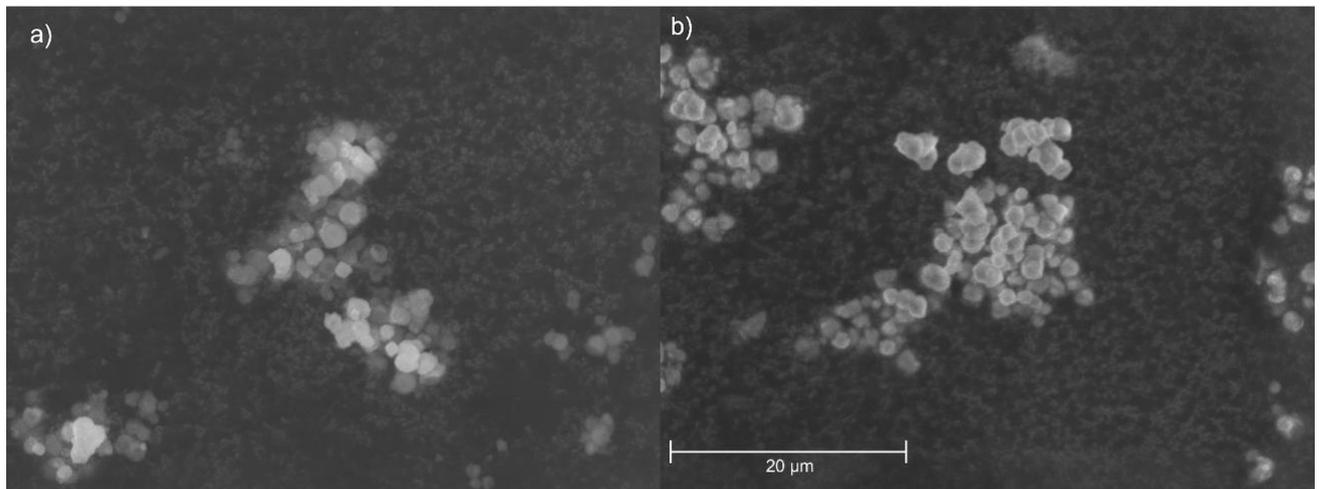

**Fig. S4**: SEM images of two different regions of $Y_{0.20}Tb_{0.80}(HCO_2)_3$, acquired with a secondary electron (SE) detector at the same magnification.

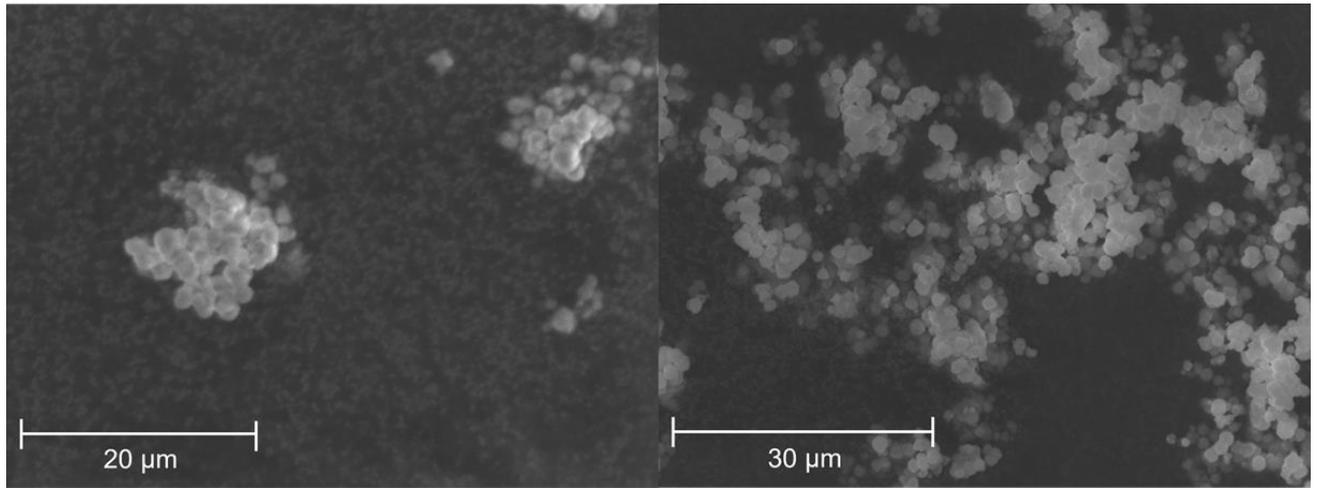

**Fig. S5**: SEM images of two different regions of $Y_{0.40}Tb_{0.60}(HCO_2)_3$, acquired with a secondary electron (SE) detector at different magnifications.

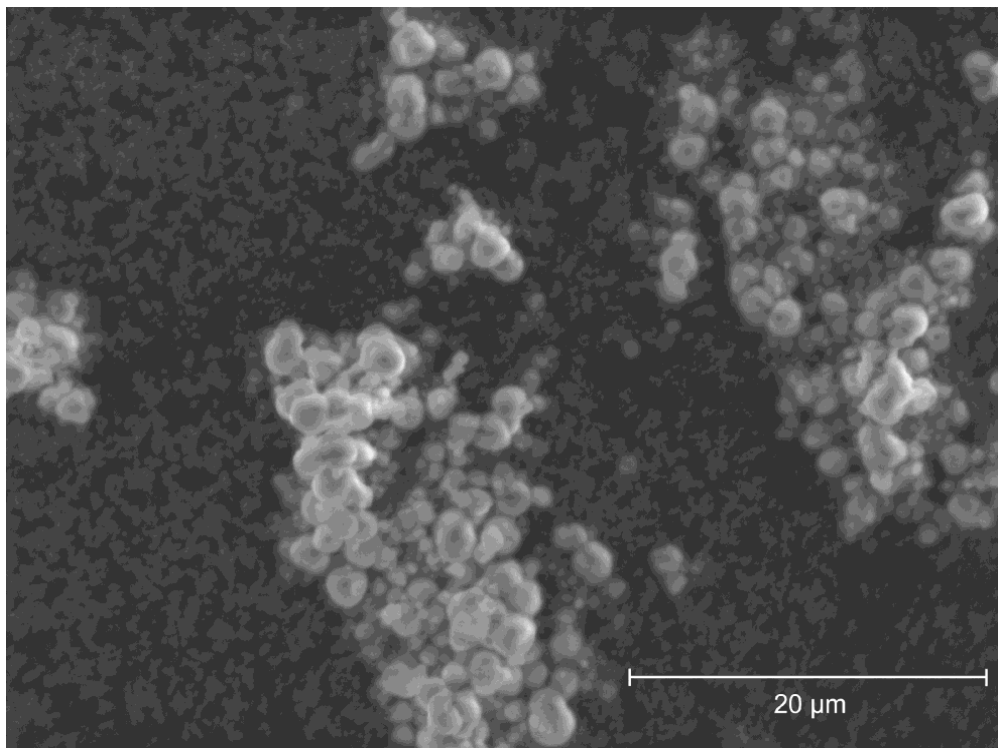

**Fig. S6**: SEM image of $Y_{0.60}Tb_{0.40}(HCO_2)_3$, acquired with a secondary electron (SE) detector.

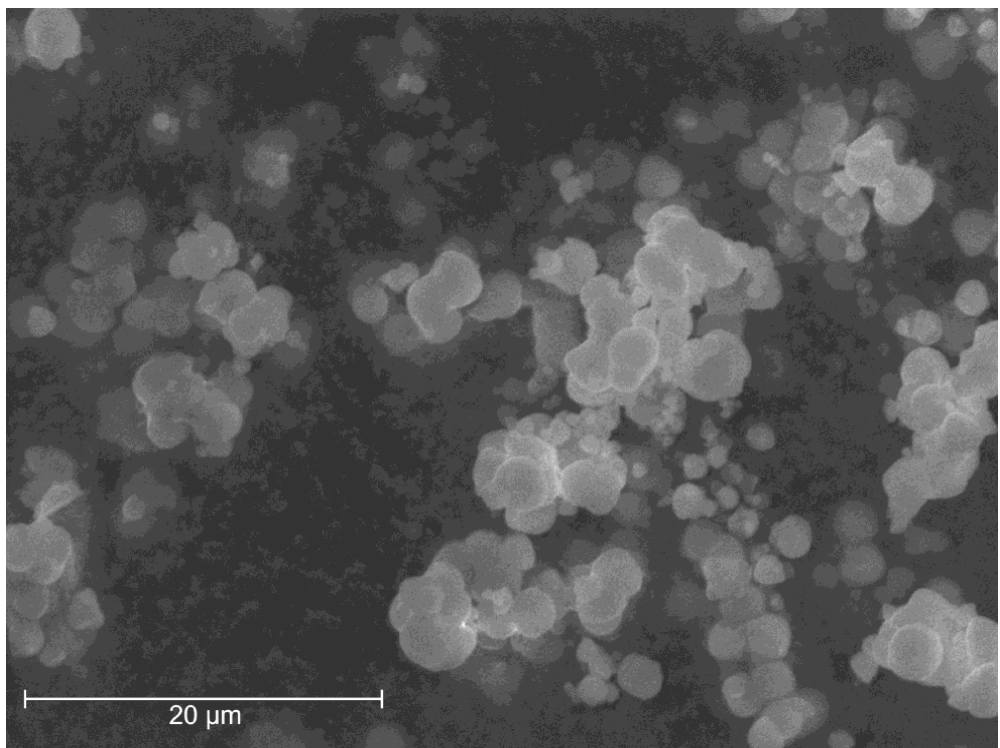

**Fig. S7**: SEM image of $Y_{0.80}Tb_{0.20}(HCO_2)_3$, acquired with a secondary electron (SE) detector.

**Table S1**. EDX atomic percentages of yttrium and terbium for $Y_{0.025}Tb_{0.975}(HCO_2)_3$ determined on different points and areas of the observed region under SEM.

| Element | Y | Tb |
|---|---|---|
| Atomic % | -1.36 | 101.36 |
| | 3.17 | 96.83 |
| | 4.86 | 95.14 |
| | 1.44 | 98.56 |
| | 1.87 | 98.13 |
| | -0.72 | 100.72 |
| | 5.64 | 94.36 |
| | 2.57 | 97.43 |
| | 2.92 | 97.08 |
| Average | 2.27 | 97.73 |
| Standard Deviation | 2.30 | 2.30 |

**Table S2**. EDX atomic percentages of yttrium and terbium for $Y_{0.05}Tb_{0.95}(HCO_2)_3$ determined on different points and areas of the observed region under SEM.

| Element | Y | Tb |
|---|---|---|
| Atomic % | 5.89 | 94.11 |
| | 1.58 | 98.42 |
| | 5.53 | 94.47 |
| | 3.54 | 96.46 |
| | 5.74 | 94.26 |
| | 7.01 | 92.99 |
| | 3.21 | 96.79 |
| | 3.55 | 96.45 |
| | 4.42 | 95.58 |
| | 5.45 | 94.55 |
| | 7.5 | 92.50 |
| Average | 4.87 | 95.14 |
| Standard Deviation | 1.77 | 1.77 |

**Table S3**. EDX atomic percentages of yttrium and terbium for $Y_{0.10}Tb_{0.90}(HCO_2)_3$ determined on different points and areas of the observed region under SEM.

| Element | Y | Tb |
|---|---|---|
| Atomic % | 17.85 | 82.15 |
| | 15.23 | 84.77 |
| | 17.84 | 82.16 |
| | 8.51 | 91.49 |
| | 30.84 | 69.16 |
| | 8.15 | 91.85 |
| | 9.78 | 90.22 |
| | 14.96 | 85.04 |
| | 17.17 | 82.83 |
| | 11.78 | 88.22 |
| | 10.87 | 89.13 |
| | 8.43 | 91.57 |
| Average | 14.28 | 85.72 |
| Standard Deviation | 6.41 | 6.41 |

**Table S4**. EDX atomic percentages of yttrium and terbium for $Y_{0.20}Tb_{0.80}(HCO_2)_3$ determined on different points and areas of the two observed regions under SEM.

| Element | Y | Tb |
|---|---|---|
| Atomic % | 28.86 | 71.14 |
| | 25.88 | 74.12 |
| | 25.79 | 74.21 |
| | 29.9 | 70.1 |
| | 26.35 | 73.65 |
| | 27.57 | 72.43 |
| | 28.93 | 71.07 |
| | 27.31 | 72.69 |
| | 27.24 | 72.76 |
| | 29.06 | 70.94 |
| | 28.05 | 71.95 |
| | 24.12 | 75.88 |
| | 18.91 | 81.09 |
| | 23.11 | 76.89 |
| | 40.13 | 59.87 |
| | 24.88 | 75.12 |
| | 27.33 | 72.67 |
| | 23.31 | 76.69 |
| | 20.99 | 79.01 |
| | 31.25 | 68.75 |
| | 23.99 | 76.01 |
| | 33.7 | 66.3 |
| | 18.99 | 81.01 |
| Average | 26.77 | 73.23 |
| Standard Deviation | 4.65 | 4.65 |

**Table S5.** EDX atomic percentages of yttrium and terbium for $Y_{0.40}Tb_{0.60}(HCO_2)_3$ determined on different points and areas of the two observed regions under SEM.

| Element | Y | Tb |
|---|---|---|
| Atomic % | 50.64 | 49.36 |
| | 54.16 | 45.84 |
| | 48.86 | 51.14 |
| | 55.9 | 44.1 |
| | 46.25 | 53.75 |
| | 25.28 | 74.72 |
| | 43.9 | 56.1 |
| | 53.45 | 46.55 |
| | 50.24 | 49.76 |
| | 43.09 | 56.91 |
| | 55.46 | 44.54 |
| | 56.43 | 43.57 |
| | 53.75 | 46.25 |
| | 59.84 | 40.16 |
| | 43.95 | 56.05 |
| | 48.84 | 51.16 |
| | 41.87 | 58.13 |
| | 51.21 | 48.79 |
| | 46.15 | 53.85 |
| | 49.95 | 50.05 |
| | 45.07 | 54.93 |
| | 46.69 | 53.31 |
| | 49.19 | 50.81 |
| | 49.89 | 50.11 |
| | 50.69 | 49.31 |
| | 47.82 | 52.18 |
| | 49.44 | 50.56 |
| | 49.43 | 50.57 |
| | 47.09 | 52.91 |
| Average | 48.78 | 51.22 |
| Standard Deviation | 6.23 | 6.23 |

**Table S6**. EDX atomic percentages of yttrium and terbium for $Y_{0.60}Tb_{0.40}(HCO_2)_3$ determined on different points and areas of the observed region under SEM.

| Element | Y | Tb |
|---|---|---|
| Atomic % | 75.56 | 24.44 |
| | 64.87 | 35.13 |
| | 67.52 | 32.48 |
| | 59.22 | 40.78 |
| | 60.42 | 39.58 |
| | 76.05 | 23.95 |
| | 73.71 | 26.29 |
| | 63.63 | 36.37 |
| | 65.06 | 34.94 |
| | 69.92 | 30.08 |
| | 66.24 | 33.76 |
| | 67.78 | 32.22 |
| | 62.63 | 37.37 |
| | 72.24 | 27.76 |
| | 70.25 | 29.75 |
| | 73.1 | 26.9 |
| | 64.3 | 35.7 |
| | 69.05 | 30.95 |
| Average | 67.86 | 32.14 |
| Standard Deviation | 5.01 | 5.01 |

**Table S7**. EDX atomic percentages of yttrium and terbium for $Y_{0.80}Tb_{0.20}(HCO_2)_3$ determined on different points and areas of the observed region under SEM.

| Element | Y | Tb |
|---|---|---|
| Atomic % | 82.49 | 17.51 |
| | 83.6 | 16.4 |
| | 82.56 | 17.44 |
| | 83.89 | 16.11 |
| | 83.08 | 16.92 |
| | 83.78 | 16.22 |
| | 83.69 | 16.31 |
| | 84.06 | 15.94 |
| | 83.64 | 16.36 |
| | 84.08 | 15.92 |
| | 84.84 | 15.16 |
| | 82.73 | 17.27 |
| | 85.4 | 14.6 |
| | 83.36 | 16.64 |
| | 83.63 | 16.37 |
| | 83.25 | 16.75 |
| | 81.54 | 18.46 |
| | 80.9 | 19.1 |
| | 82.85 | 17.15 |
| | 83.22 | 16.78 |
| Average | 83.33 | 16.67 |
| Standard Deviation | 1.02 | 1.02 |

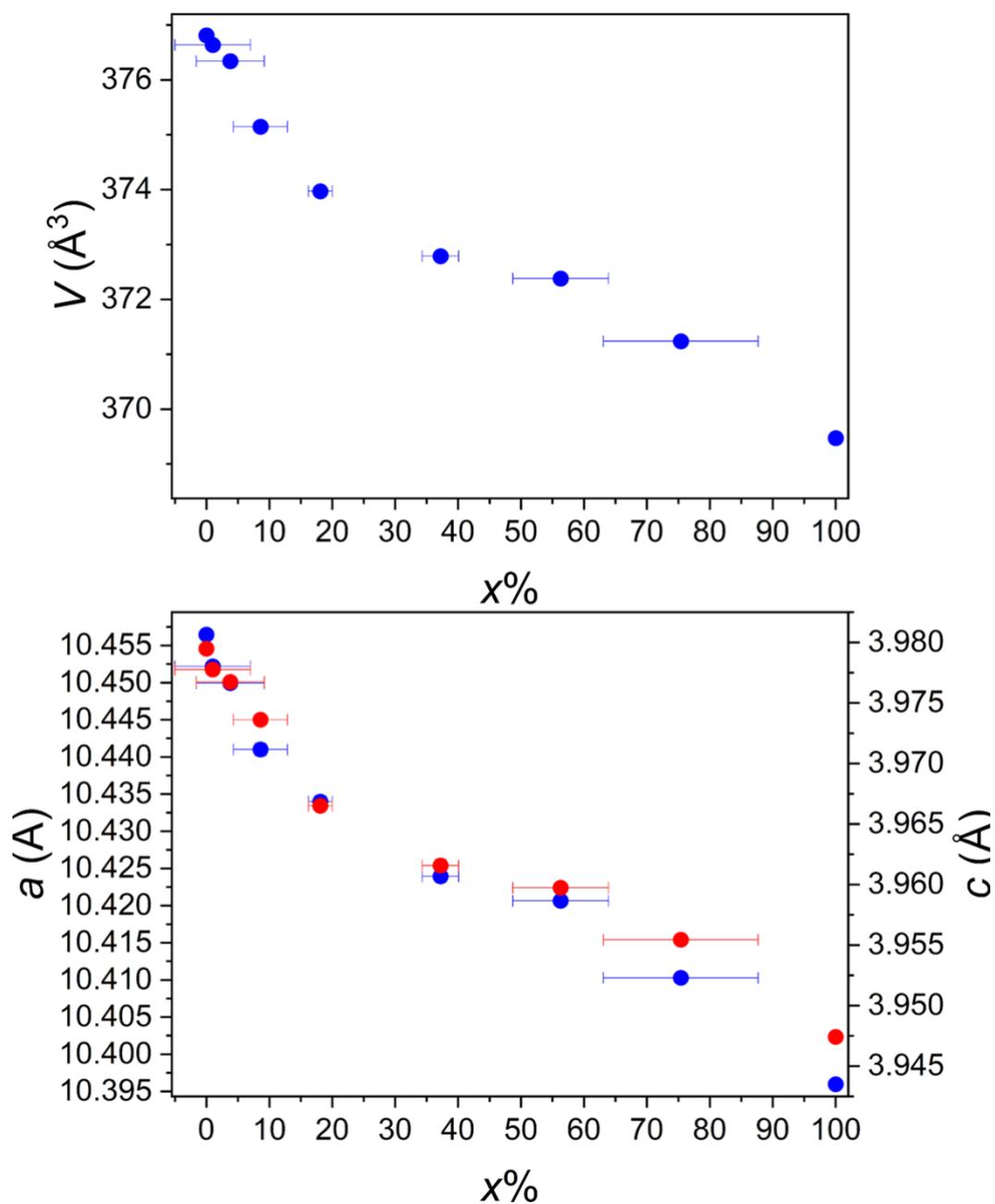

**Fig. S8** Plots of the unit cell volume (top) and lattice parameters *a* and *c* (bottom), the latter shown in blue and red, respectively, for the orthorhombic unit cell of Y$_x$Tb$_{1-x}$(HCO$_2$)$_3$. Horizontal error bars show the uncertainty on the composition determined *via* X-ray fluorescence for the solid solutions.

**Table S8:** Refinements statistic for the Le Bail refinements of powder X-ray diffraction patterns of $Y_xTb_{1-x}(HCO_2)_3$.

| Sample | $R_p$ | $R_{wp}$ | $\chi^2$ |
|---|---|---|---|
| $Tb(HCO_2)_3$ | 2.18 | 2.84 | 1.91 |
| $Y_{0.025}Tb_{0.975}(HCO_2)_3$ | 2.17 | 2.80 | 1.58 |
| $Y_{0.05}Tb_{0.95}(HCO_2)_3$ | 3.65 | 4.65 | 1.30 |
| $Y_{0.10}Tb_{0.90}(HCO_2)_3$ | 4.07 | 5.16 | 1.26 |
| $Y_{0.20}Tb_{0.80}(HCO_2)_3$ | 2.30 | 2.94 | 1.31 |
| $Y_{0.40}Tb_{0.60}(HCO_2)_3$ | 2.46 | 3.12 | 1.17 |
| $Y_{0.60}Tb_{0.40}(HCO_2)_3$ | 2.48 | 3.17 | 1.32 |
| $Y_{0.80}Tb_{0.20}(HCO_2)_3$ | 2.24 | 2.83 | 1.41 |
| $Y(HCO_2)_3$ | 2.25 | 3.03 | 2.73 |

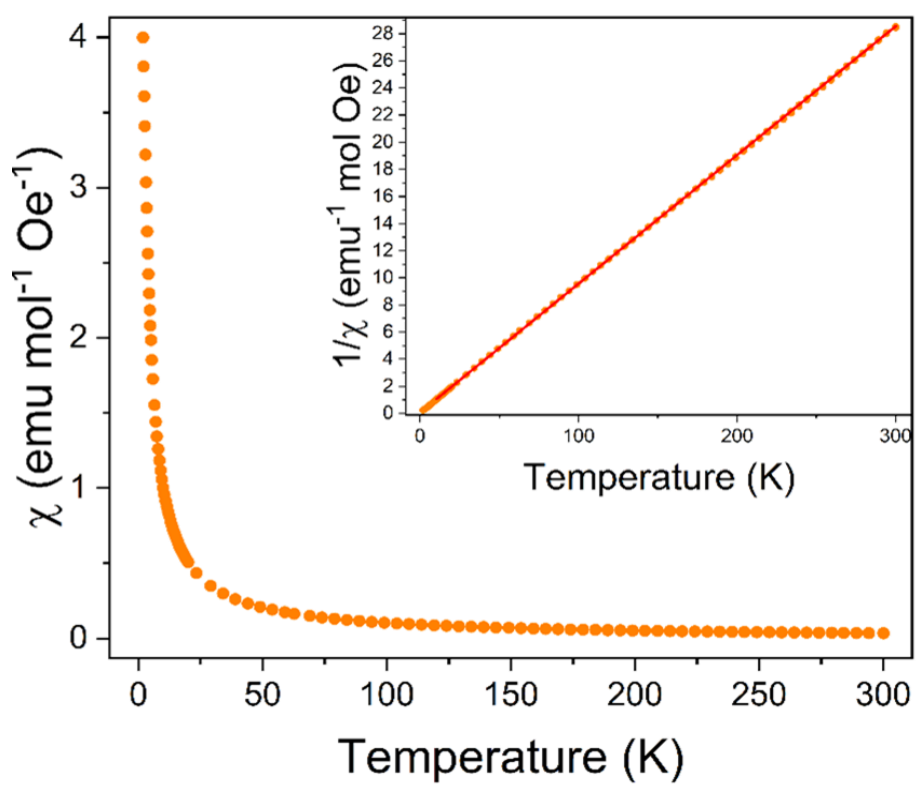

**Fig. S9**: FC susceptibility measurement for $Y_{0.025}Tb_{0.975}(HCO_2)_3$ in a 0.1 T applied magnetic field. The insert shows the inverse susceptibility and the Curie-Weiss fit to the data obtained between 10 and 300 K.

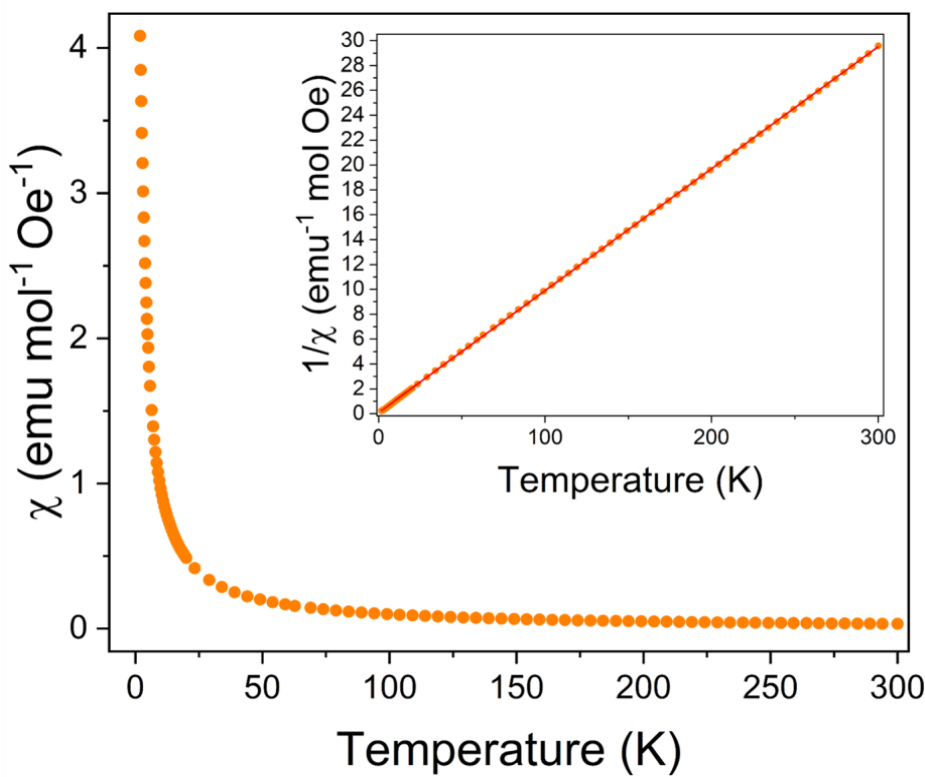

**Fig. S10**: FC susceptibility measurement for $Y_{0.05}Tb_{0.95}(HCO_2)_3$ in a 0.1 T applied magnetic field. The insert shows the inverse susceptibility and the Curie-Weiss fit to the data obtained between 10 and 300 K.

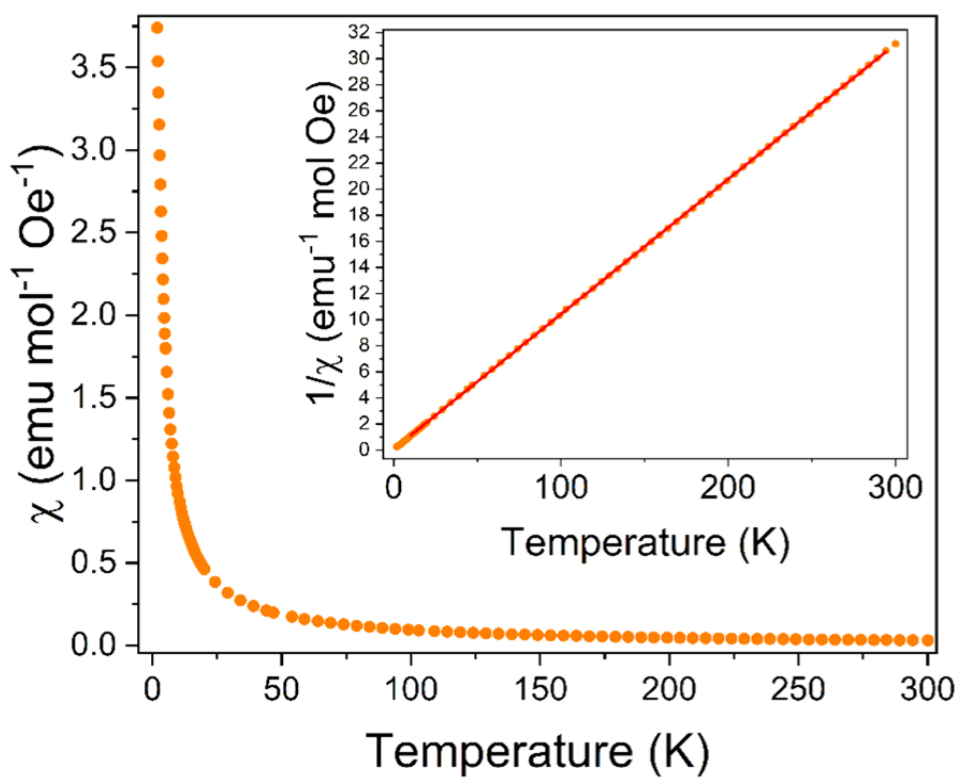

**Fig. S11**: FC susceptibility measurement for $Y_{0.10}Tb_{0.90}(HCO_2)_3$ in a 0.1 T applied magnetic field. The insert shows the inverse susceptibility and the Curie-Weiss fit to the data obtained between 10 and 300 K.

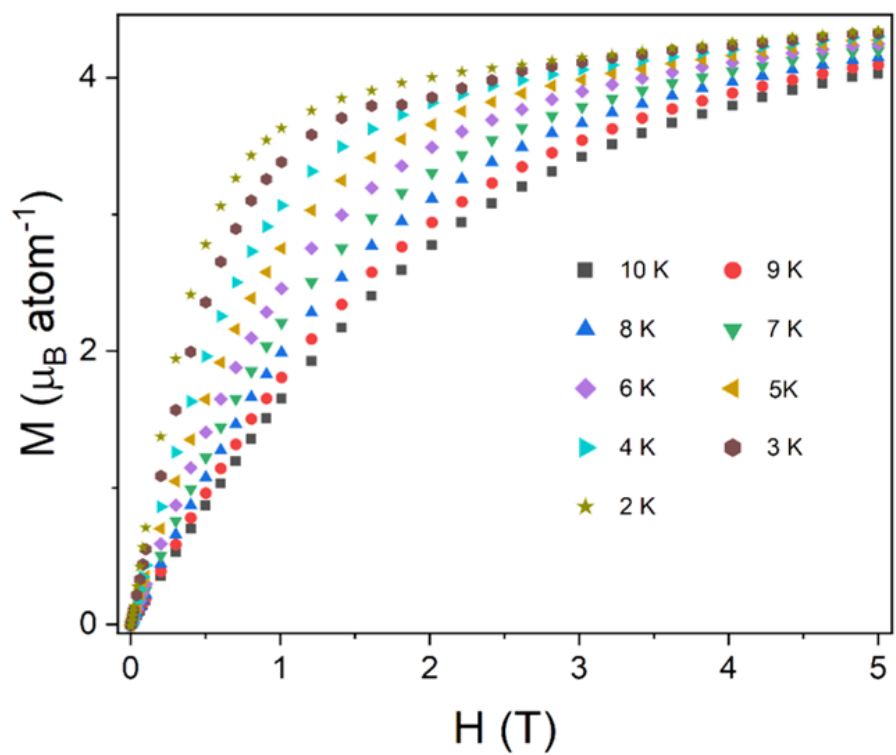

**Fig. S12**: Isothermal magnetisation measurements as a function of the applied magnetic field for $Y_{0.025}Tb_{0.975}(HCO_2)_3$.

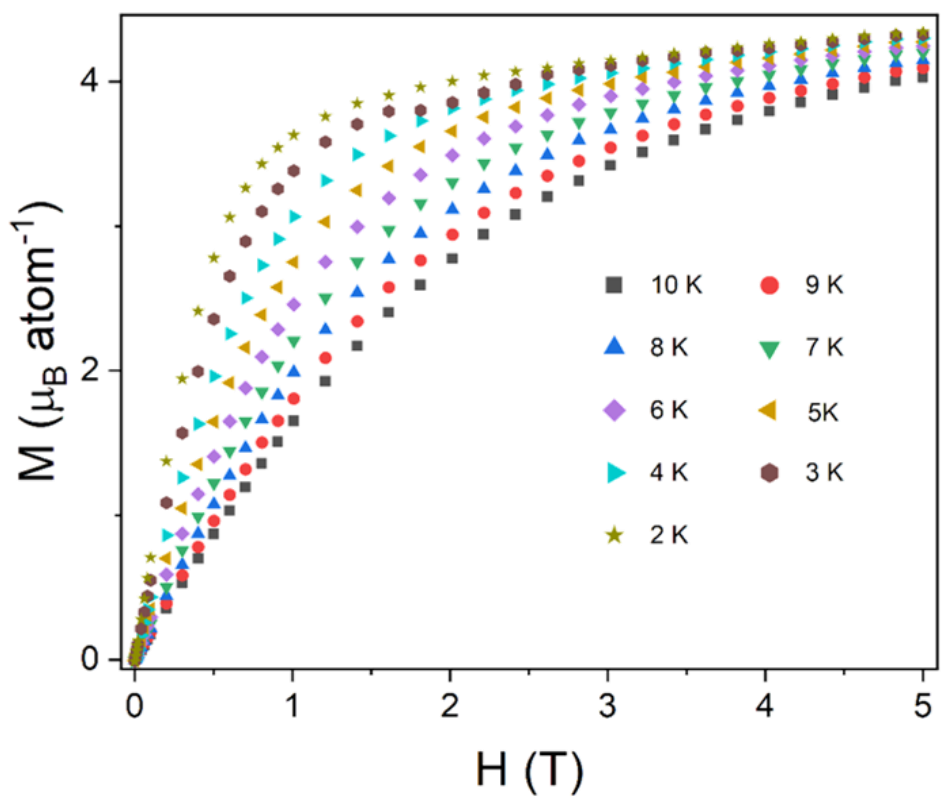

**Fig. S13**: Isothermal magnetisation measurements as a function of the applied magnetic field for $Y_{0.05}Tb_{0.95}(HCO_2)_3$.

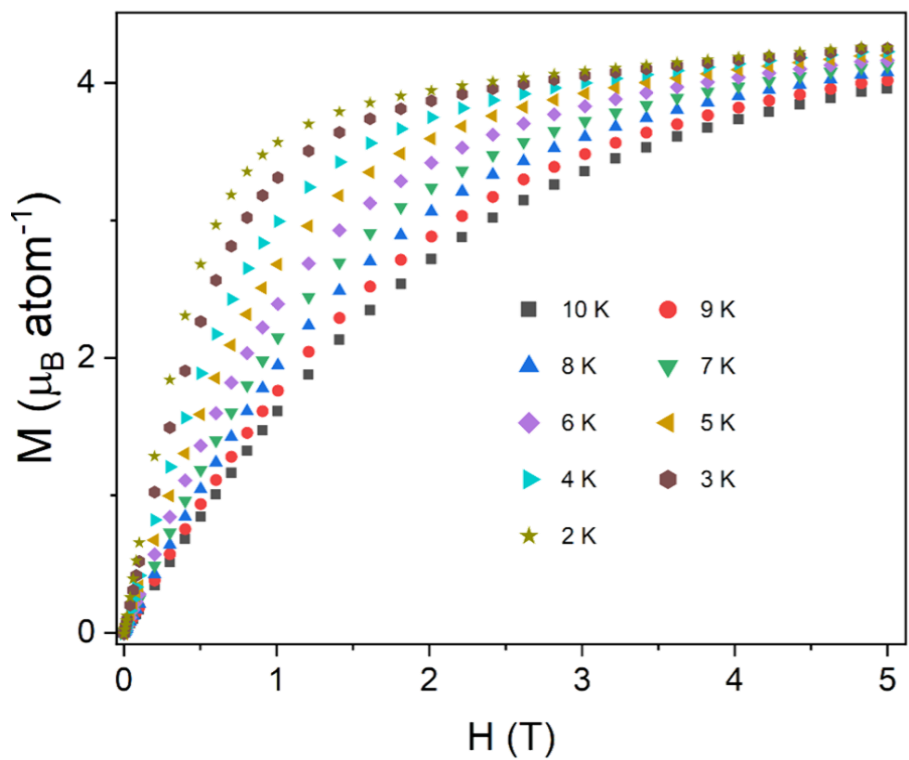

**Fig. S14**: Isothermal magnetisation measurements as a function of the applied magnetic field for $Y_{0.10}Tb_{0.90}(HCO_2)_3$.

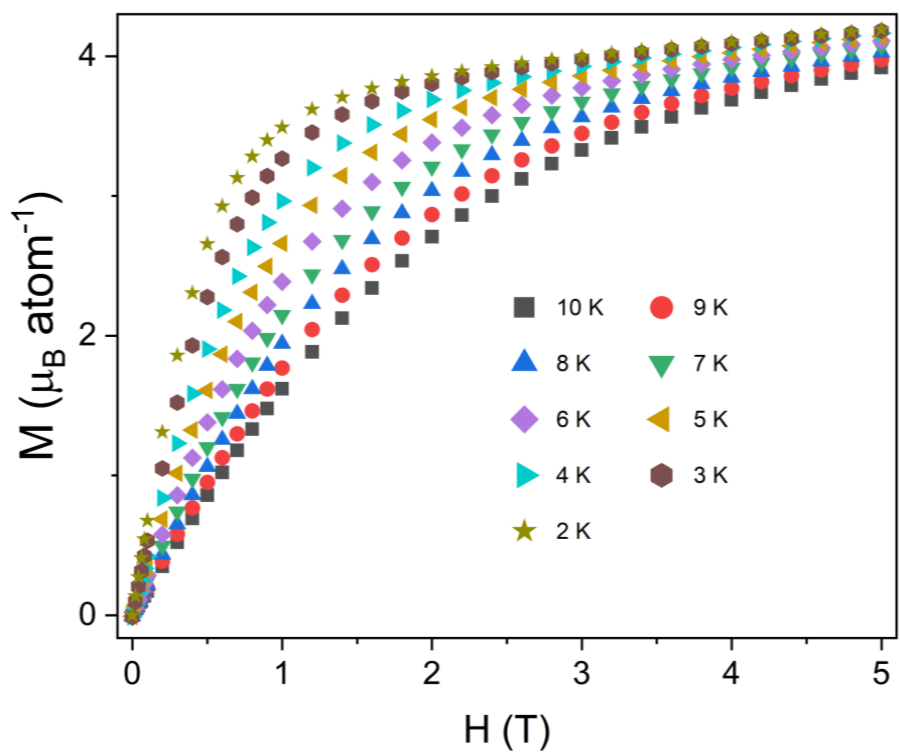

**Fig. S15**: Isothermal magnetisation measurements as a function of the applied magnetic field for $Y_{0.20}Tb_{0.80}(HCO_2)_3$.

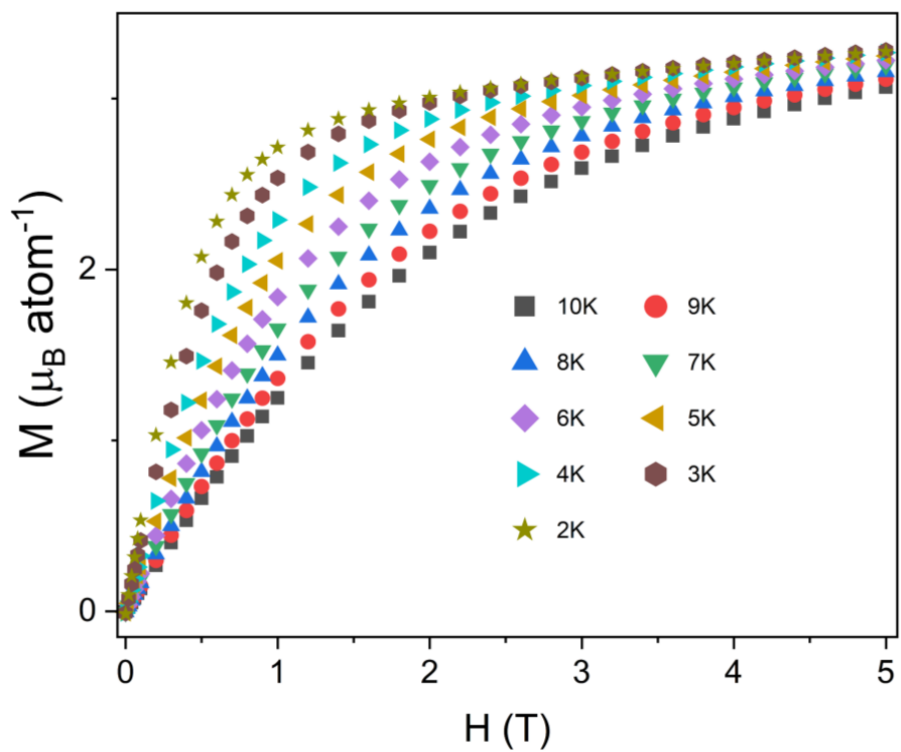

**Fig. S16**: Isothermal magnetisation measurements as a function of the applied magnetic field for $Y_{0.40}Tb_{0.60}(HCO_2)_3$.

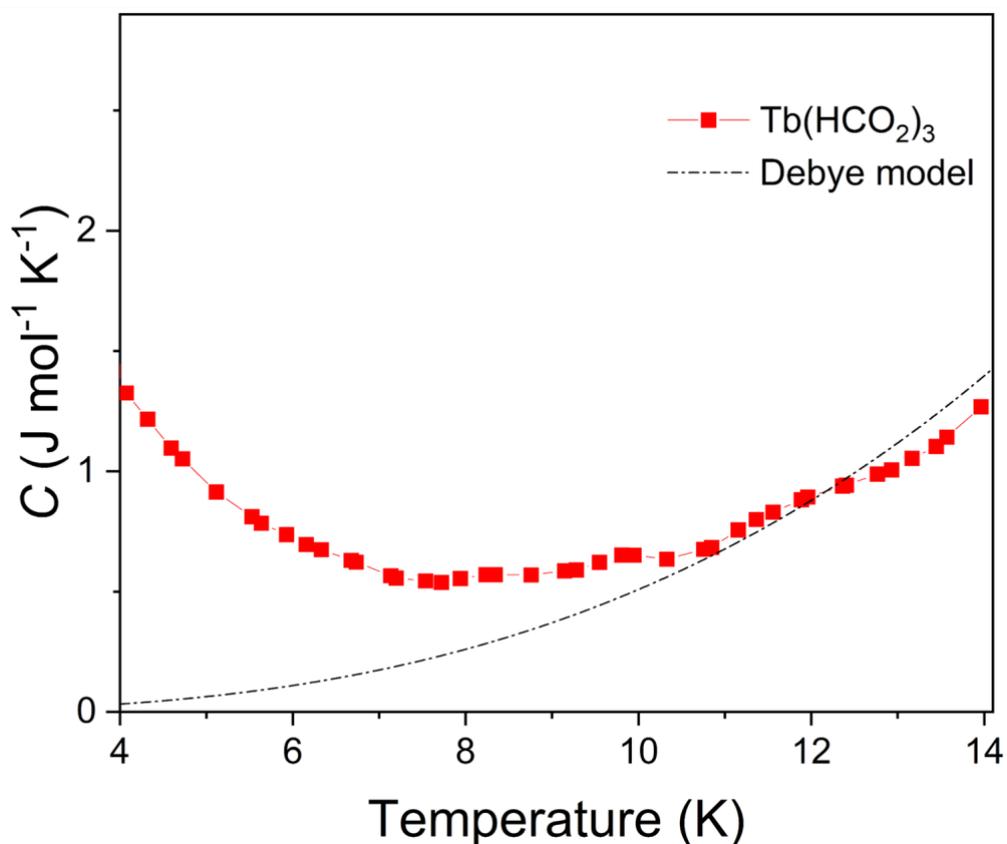

**Fig. S17:** Heat capacity data for $Tb(HCO_2)_3$ between 4 to 14 K and Debye function plotted in the whole range of the plot, with the Debye temperature determined fitting the data between 8 to 14 K.

**Table S9:** Hyperfine coupling Δ parameter for other Tb-containing systems.

| System | Oxidation state | Δ (K) | Ref |
|---|---|---|---|
| Tb-metal | 0 | 0.45 | 25 |
| Tb@SrF$_2$ | 3 | 0.476 | 26 |
| Tb(CF$_3$SO$_3$)$_3$.$_9$H$_2$O | 3 | 0.470 | 27 |

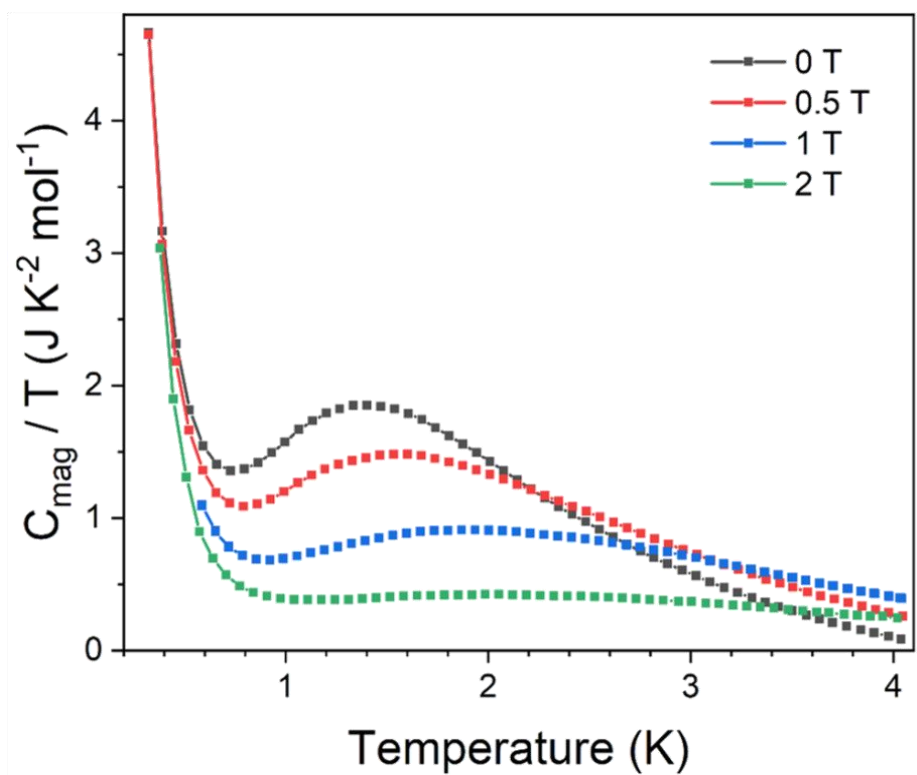

**Fig. S18**: Normalised magnetic heat capacity $C_{mag}/T$ as a function of temperature $T$ for Y$_{0.10}$Tb$_{0.90}$(HCO$_2$)$_3$.

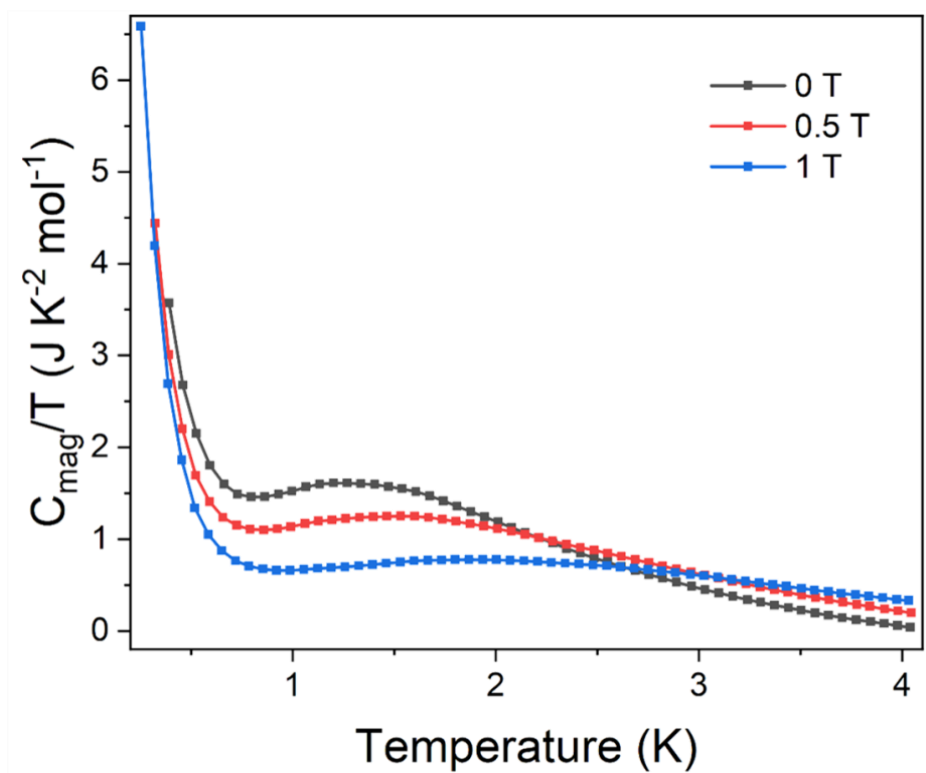

**Fig. S19**: Normalised magnetic heat capacity $C_{mag}/T$ as a function of temperature $T$ for $Y_{0.20}Tb_{0.80}(HCO_2)_3$.

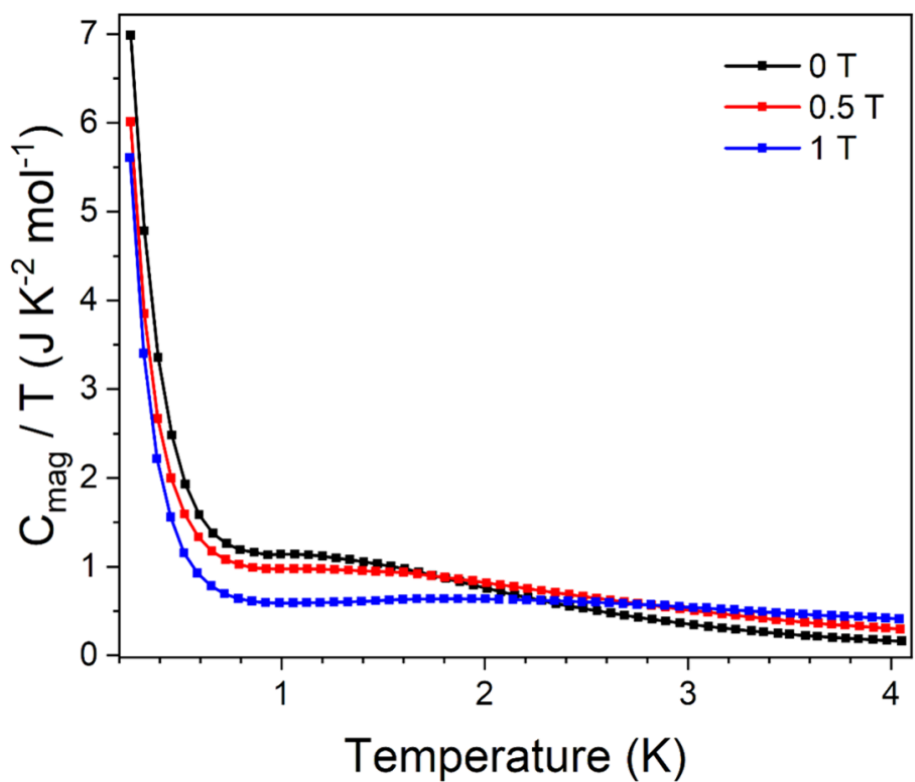

**Fig. S20**: Normalised magnetic heat capacity $C_{mag}/T$ as a function of temperature $T$ for $Y_{0.40}Tb_{0.60}(HCO_2)$.

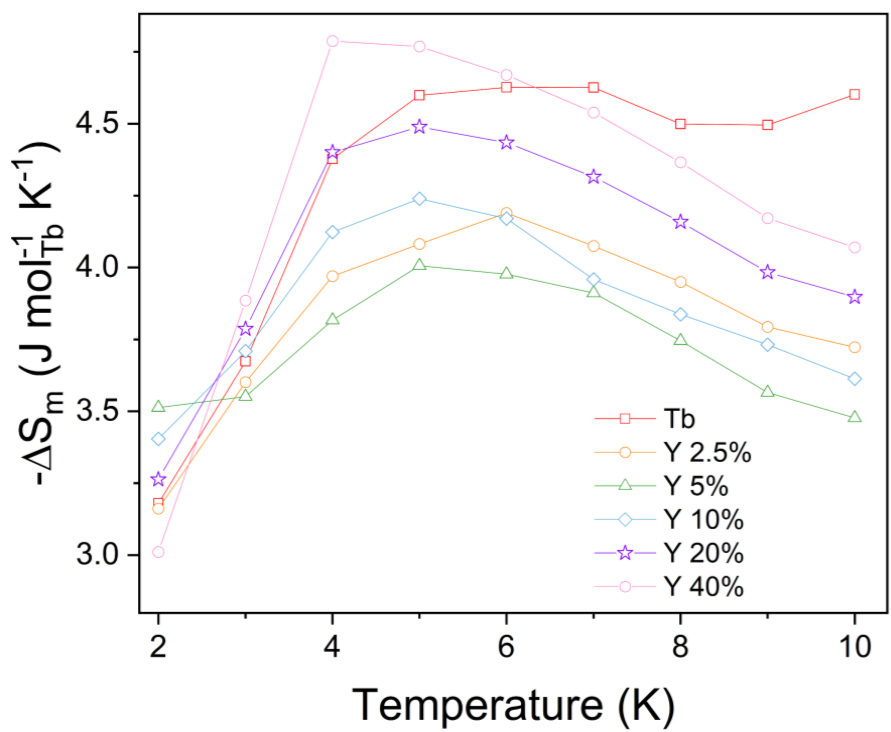

**Fig. S21**: Molar magnetic entropy changes, normalised per amount of Tb, extracted from the magnetisation data for Tb(HCO$_2$)$_3$ and the Tb$_{1-x}$Y$_x$(HCO$_2$)$_3$ (x = 0.025, 0.05, 0.10, 0.20, 0.40) solid solutions for ΔB = 5-0T.

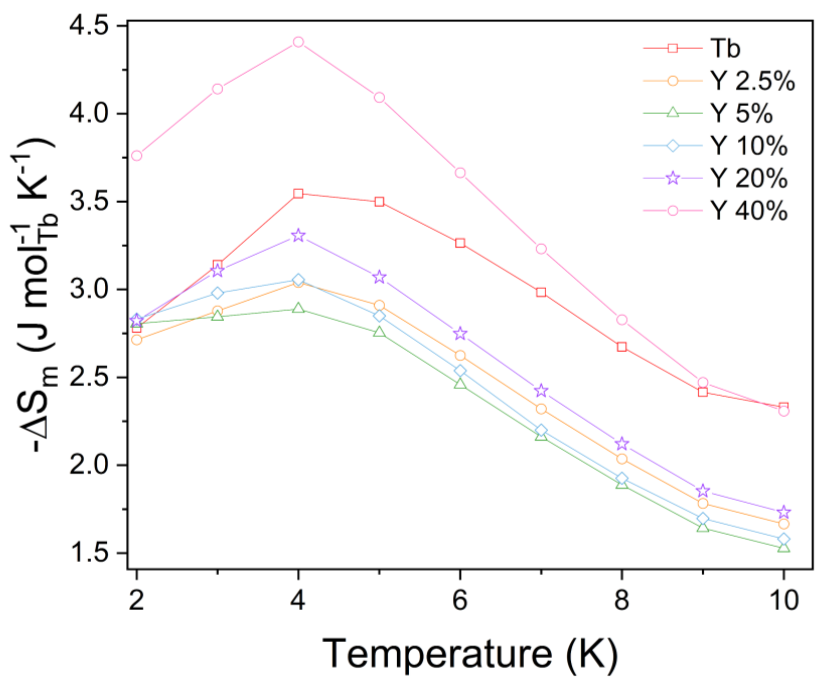

**Fig. S22**: Molar magnetic entropy changes, normalised per amount of Tb, extracted from the magnetisation data for Tb(HCO$_2$)$_3$ and the Tb$_{1-x}$Y$_x$(HCO$_2$)$_3$ (x = 0.025, 0.05, 0.10, 0.20, 0.40) solid solutions for ΔB = 2-0T.

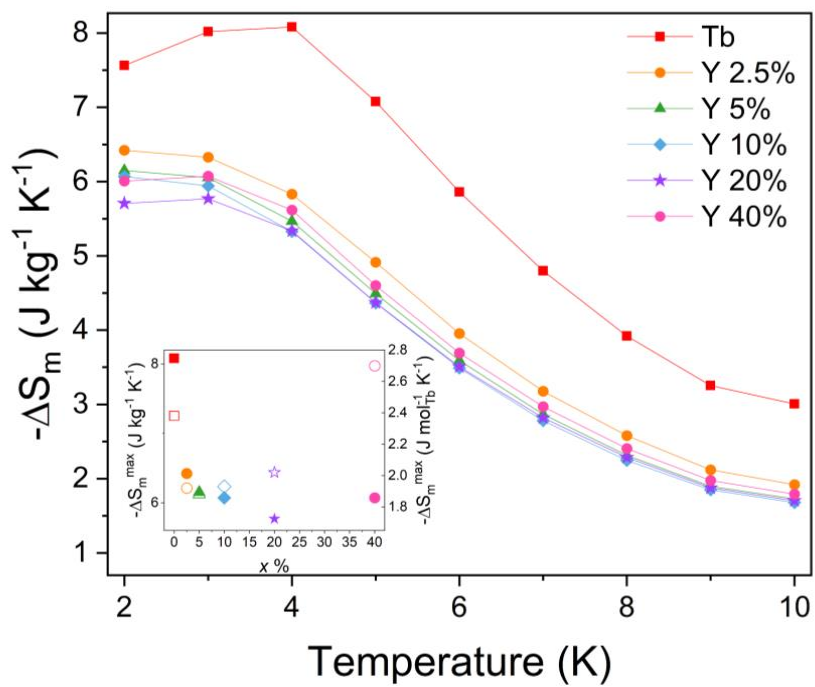

**Fig. S23**: Gravimetric magnetic entropy changes extracted from the magnetisation data for $Tb(HCO_2)_3$ and the $Tb_{1-x}Y_x(HCO_2)_3$ (x = 0.025, 0.05, 0.10, 0.20, 0.40) solid solutions for $\Delta B$ = 1-0 T. The inset shows the values of $-\Delta S_m^{max}$ in gravimetric (full symbols) and normalised molar (hollow symbols) units for each sample.

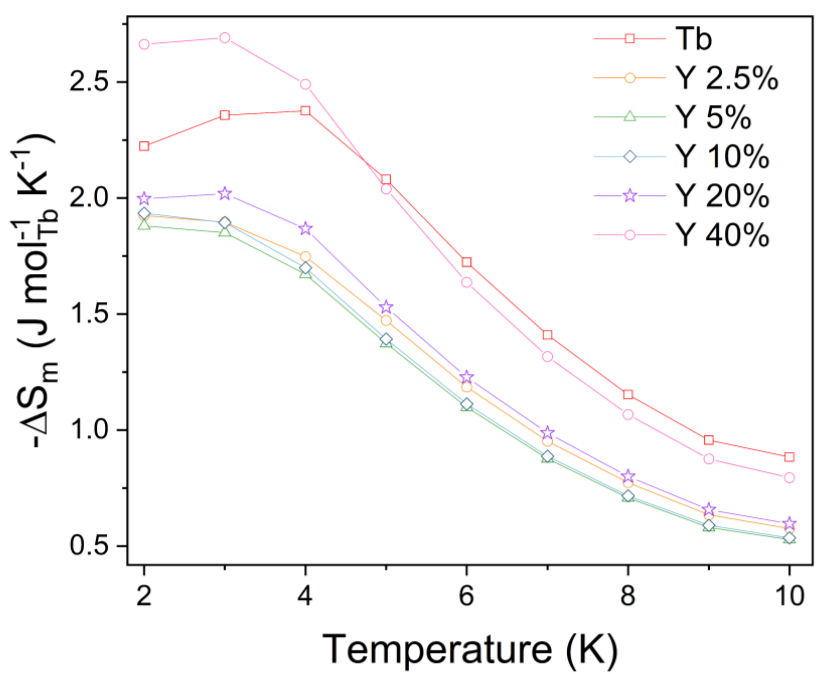

**Fig. S24**: Molar magnetic entropy changes, normalised per amount of Tb, extracted from the magnetisation data for $Tb(HCO_2)_3$ and the $Tb_{1-x}Y_x(HCO_2)_3$ (x = 0.025, 0.05, 0.10, 0.20, 0.40) solid solutions for $\Delta B$ = 1-0T.

**Table S10:** Density values $\delta$ for $Tb_{1-x}Y_x(HCO_2)_3$ with $x$ = 0.025, 0.05, 0.10, 0.20 and 0.40, respectively.

| x | 0.025 | 0.05 | 0.10 | 0.20 | 0.40 |
|---|---|---|---|---|---|
| $\delta$ (g cm$^{-3}$) | 3.86 | 3.84 | 3.81 | 3.73 | 3.55 |

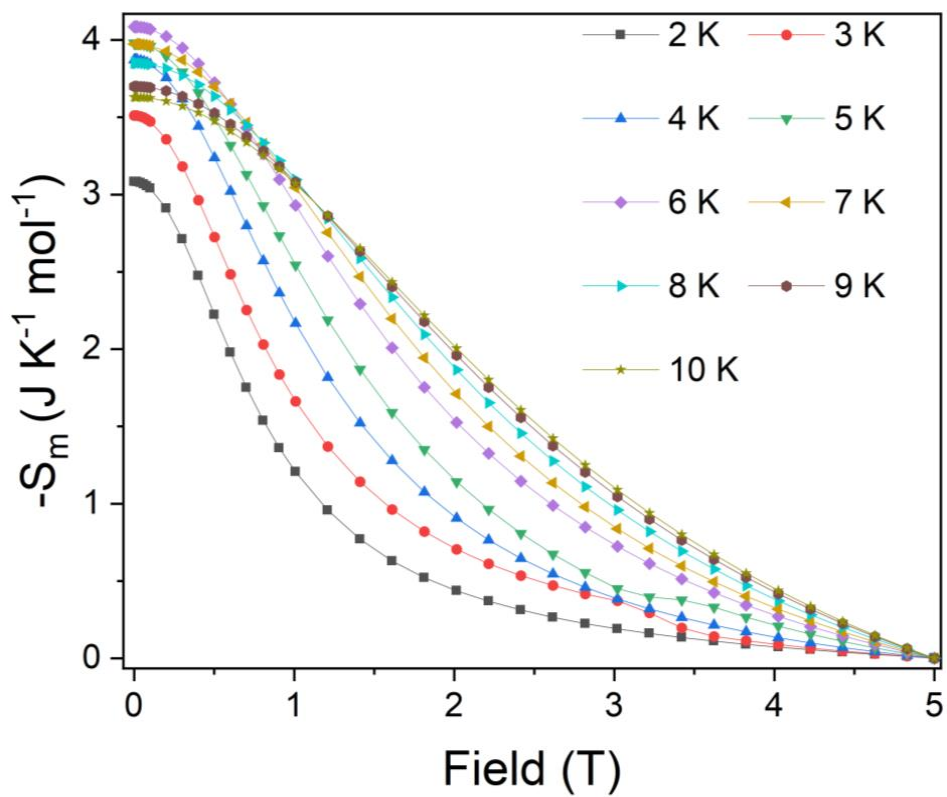

**Fig. S25**: Variation of magnetic entropy as a function of applied magnetic field extracted from the magnetisation data for $Tb_{0.975}Y_{0.025}(HCO_2)_3$ at different temperatures.

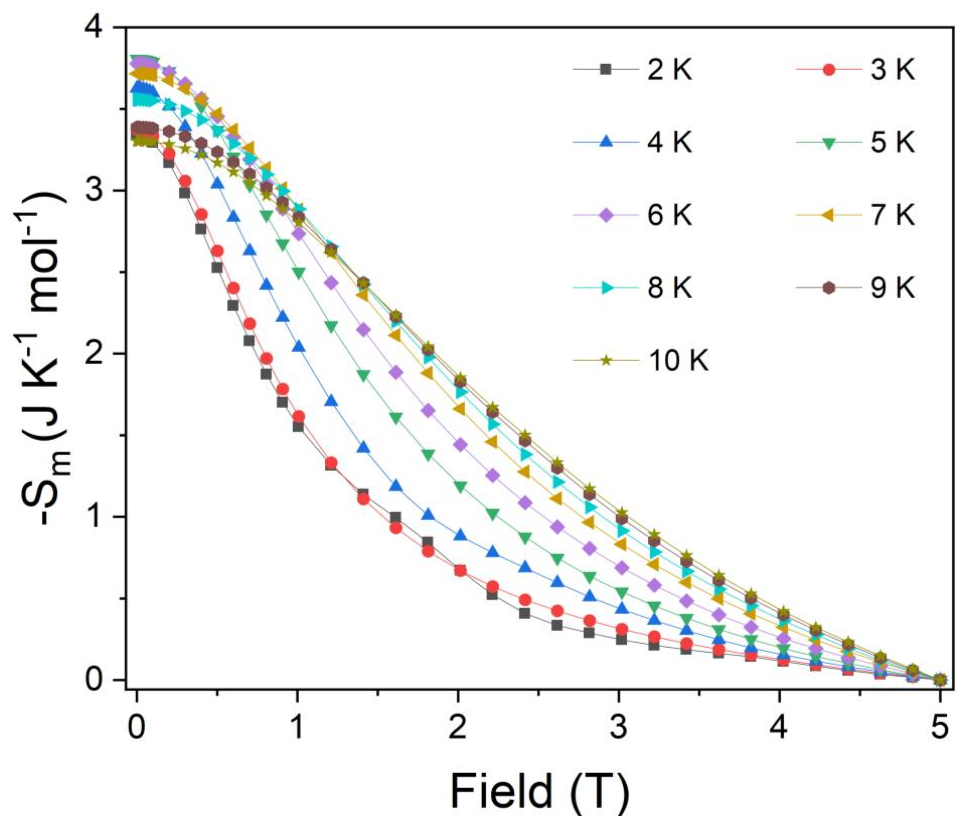

**Fig. S26**: Variation of magnetic entropy as a function of applied magnetic field extracted from the magnetisation data for Tb$_{0.95}$Y$_{0.05}$(HCO$_2$)$_3$ at different temperatures.

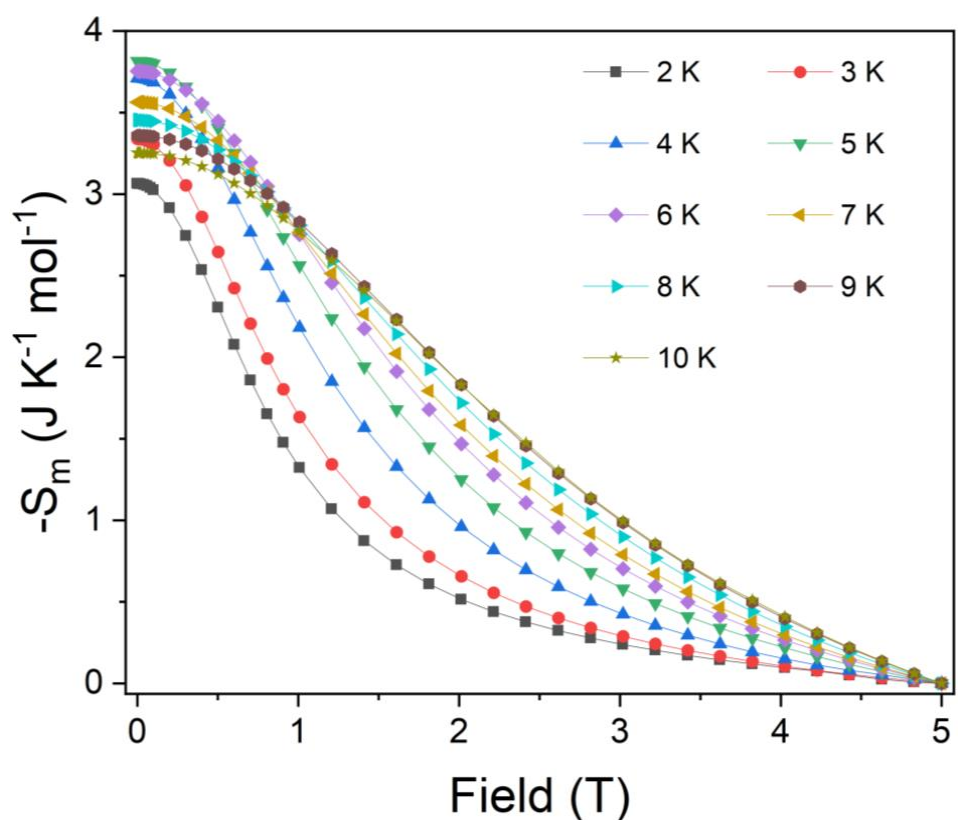

**Fig. S27**: Variation of magnetic entropy as a function of applied magnetic field extracted from the magnetisation data for Tb$_{0.90}$Y$_{0.10}$(HCO$_2$)$_3$ at different temperatures.

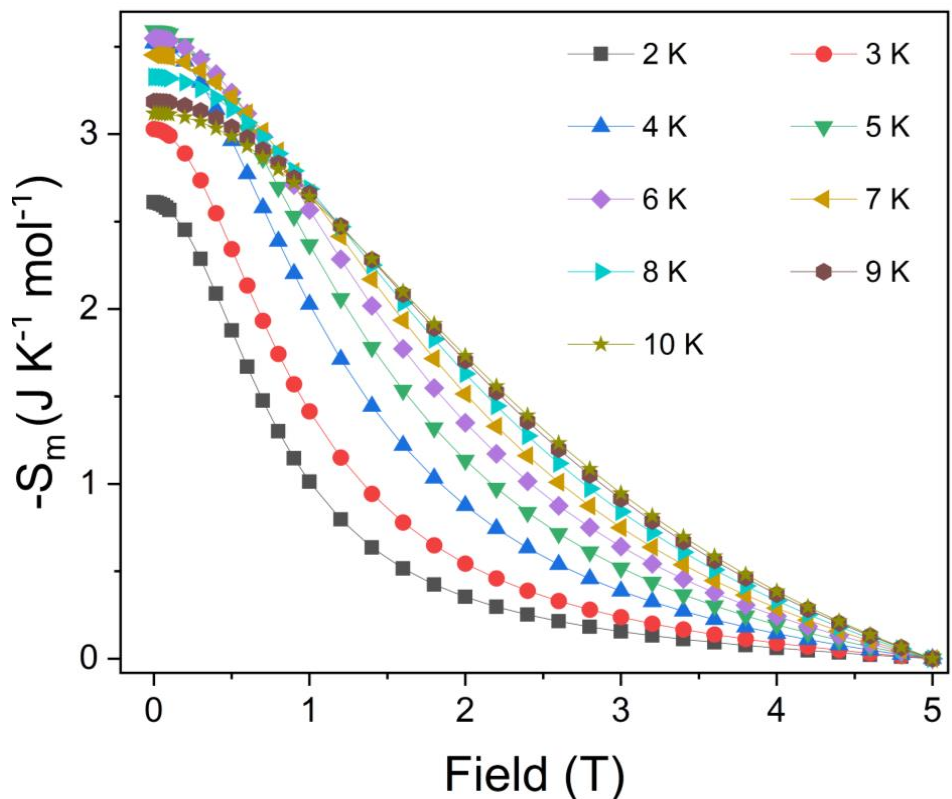

**Fig. S28**: Variation of magnetic entropy as a function of applied magnetic field extracted from the magnetisation data for Tb$_{0.80}$Y$_{0.20}$(HCO$_2$)$_3$ at different temperatures.

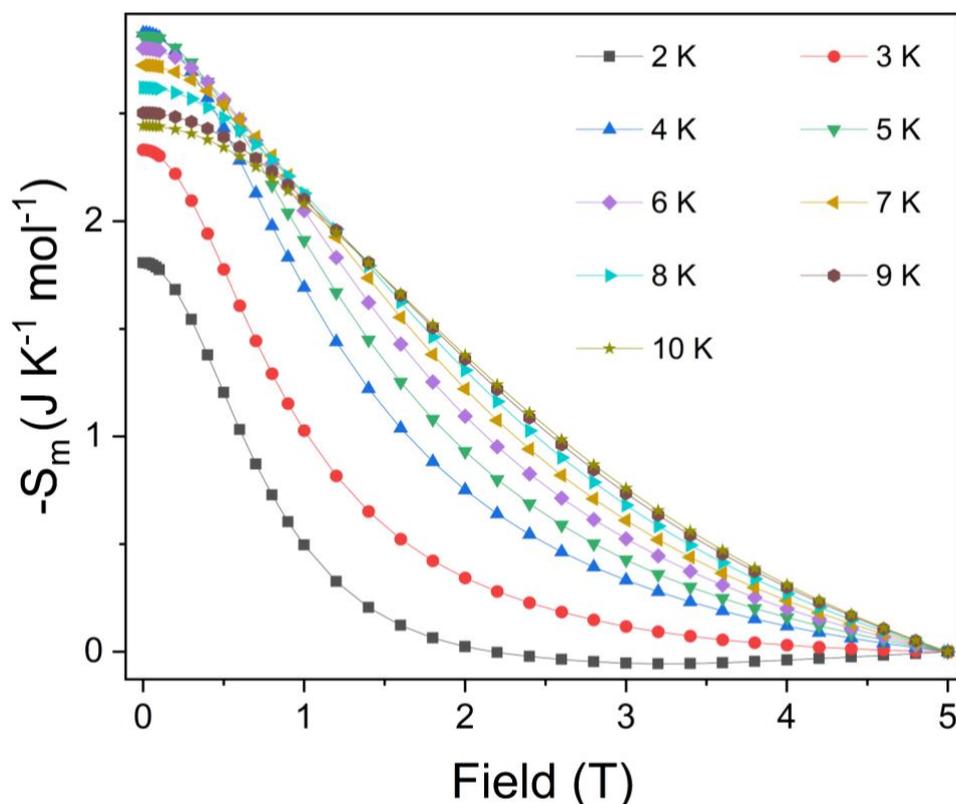

**Fig. S29**: Variation of magnetic entropy as a function of applied magnetic field extracted from the magnetisation data for Tb$_{0.60}$Y$_{0.40}$(HCO$_2$)$_3$ at different temperatures.